\documentclass[12pt]{article}
\pdfoutput=1
\usepackage[a4paper]{geometry}
 
\usepackage[T1]{fontenc} 
\usepackage{comment}

\usepackage{jheppub,hyperref,float,array,adjustbox,mathtools,physics, xcolor}

\usepackage{xcolor,tikz,pgfplots,amsmath,amsfonts}
\usepackage{graphicx}
\usetikzlibrary{matrix,calc,positioning,decorations.markings,decorations.pathmorphing,decorations.pathreplacing}
\usetikzlibrary{arrows,cd}
\usetikzlibrary{shapes.misc}
\usetikzlibrary {shapes.geometric}
\usetikzlibrary {shapes.multipart}
\usetikzlibrary{decorations.markings}
\usetikzlibrary{backgrounds}
\usepackage{bbding}

\usepackage{tikz}
\usetikzlibrary{automata,positioning,calc}
\usetikzlibrary{decorations.markings}
\tikzset{->-/.style={decoration={markings, mark=at position #1 with {\arrow{>}}},postaction={decorate}}}
\tikzset{-<-/.style={decoration={markings, mark=at position #1 with {\arrow{<}}},postaction={decorate}}}
\tikzset{auto shift/.style={auto=right,->, to path={ let \p1=(\tikztostart),\p2=(\tikztotarget), \n1={atan2(\y2-\y1,\x2-\x1)},\n2={\n1+180} in ($(\tikztostart.{\n1})!1mm!270:(\tikztotarget.{\n2})$) -- ($(\tikztotarget.{\n2})!1mm!90:(\tikztostart.{\n1})$) \tikztonodes}}}

\usetikzlibrary {shapes.geometric}

\usepackage{float}

\title{
\begin{center}
A new 4d $\mathcal{N}=1$ duality from the superconformal index
\end{center}
}

\author[a]{Antonio Amariti,}	
\author[b]{Fabio Mantegazza,}

\affiliation[a]{INFN, Sezione di Milano, Via Celoria 16, I-20133 Milano, Italy}
\affiliation[b]{Dipartimento di Fisica, Università degli studi di Milano, Via Celoria 16, I-20133, Milano, Italy}

\emailAdd{antonio.amariti@mi.infn.it}
\emailAdd{fabiomanto99@gmail.com}

\abstract{
 In this paper we propose a physical derivation of a 4d conjectural duality for $USp(2N)$ with an   anti-symmetric rank-two tensor and fundamental flavors,
 in presence of a non-trivial superpotential. 
This duality has been conjectured as a consequence  of an exact identity between the superconformal indices of the two phases, proved in the mathematical literature.
Here we show that the duality can be derived 
by a combined sequence of known dualities, deconfinement of tensor matter, RG flow and Higgsing. Furthermore, by following these steps on the superconformal index, we provide an alternative derivation of the integral identity as well.
 }

\begin{document}

\maketitle
\flushbottom
\allowdisplaybreaks 

\section{Introduction}

The existence of equivalent descriptions of a given physical system in terms of  quantum field theories with distinct degrees of freedom
is a broad definition of duality.
Supersymmetry is a fertile playground to look for such dualities, because many non-perturbative aspects are under a satisfying control.  
In this way it is possible to look for infrared dualities relating weakly coupled phases to strongly coupled ones, generalizing the concept 
of electromagnetic duality to non-abelian gauge theories.

The pivotal example for 4d $\mathcal{N}=1$ theories  was originally found in \cite{Seiberg:1994pq} for $SU(N)$ SQCD and it is referred to as Seiberg duality. Various generalizations 
of such duality have been then deeply investigated. Such generalizations involve other gauge groups, matter fields in various representations, 
different dimensionalities and different degrees of supersymmetry.

Depending on the type of duality many checks are possible. For example one can compare the moduli space, the global symmetries and the anomalies. It is also possible to
deform the theories by adding relevant operators and then study the robustness of the duality.
 Such arguments are necessary conditions for the validity of the dualities, they are quite strong and in many cases the conjectured dualities are believed to 
represent equivalent effective descriptions of the same physical system.

Further checks have been performed in the last decades thanks to the application of the exact techniques originating from supersymmetric localization on curved space.
In the physical language such checks correspond to  the matching of supersymmetric partition functions on curved backgrounds
between the dual phases. 
Depending on the background these partition functions are in general formulated in terms of matrix integrals of special functions  (e.g. Gamma functions)
and the results of such integrations often correspond  to a counting problem of specific sets of protected operators
of the theory on flat space. Hence the non trivial matching of these matrix integrals is a quite robust check of the conjectured dualities.
This last aspect is crucial not only as a check of the dualities but especially as a starting point for the search of new dualities.

A first step in this direction consists of looking for exact identities in the mathematical literature of matrix integrals. When the matrix integrals have a clear physical interpretation 
it is natural to conjecture that a physical duality follows from an integral identity.

This line of thoughts led in the last years to propose large classes of new dualities. Even if reading a physical duality from a mathematical identity is quite fascinating 
a more compelling analysis consists of deriving such a “new” duality in a physical language.

This last steps is not \emph{per se} mandatory, indeed it may be possible that some of these “new” dualities are genuinely new, but in many cases it is possible to show that they can be derived by applying other known dualities (e.g. the original Seiberg duality for $SU(N)$ SQCD and/or its generalization to  $USp(2N)$ SQCD due to Intriligator and Pouliot \cite{Intriligator:1995ne}) in addition to other techniques,  as the Berkooz deconfinement \cite{Berkooz:1995km,Luty:1996cg}, RG flows and the Higgs mechanism.
Similar construction recently appeared in the literature (see for example \cite{Benvenuti:2020gvy,Benvenuti:2021nwt,Comi:2022aqo,Bajeot:2022kwt,Amariti:2022wae,Bottini:2022vpy,Bajeot:2022lah,Bajeot:2023gyl,Amariti:2023wts}).

An avatar of this idea has been established in \cite{Bajeot:2022kwt}, where it was shown that thanks to such a procedure the confining dualities classified in \cite{Csaki:1996sm,Csaki:1996zb} could be derived using only 
Seiberg and  Intriligator-Pouliot dualities.

In this paper we focus on an identity originally conjectured in \cite{rains2012elliptic} and then proven in \cite{vandebult2009elliptic}. Such identity relates hypergeometric elliptic integrals and it corresponds 
to the matching of the supersymmetric partition functions of 4d $\mathcal{N}=1$  gauge theories on $S^3 \times S^1$. Such partition functions, in the superconformal case, are associated to the counting of semi-short chiral operators of the SCFT and they correspond  (up to a phase associated to the supersymmetric Casimir energy \cite{Assel:2015nca})  to the superconformal index \cite{Kinney:2005ej,Romelsberger:2005eg}. In general the superconformality is not strictly necessary for  the definitions of such objects, and one can refer either to the partition function on the curved background or to the supersymmetric index. Here we will keep the last notion.
In the physical language the duality that can be read from the identity of \cite{rains2012elliptic,vandebult2009elliptic} has been proposed in \cite{Spiridonov:2009za}.  
The electric phase consists of an $USp(2M)$ gauge theory with an antysimmetric $A$ and fundamentals. Among the fundamentals there are four fundamentals $W$ charged under an $SU(4)$ gauge symmetry that do not interact in any superpotential deformations. The other fundamentals $Q_i$ are charged under a $\prod_{i=1}^{k} USp(2l_i)$
flavor symmetry, and they interact with the antisymmetric through a superpotential $W = \sum_{i=1}^{k} A^{n_i} Q_i^2$ where $n_i\neq n_j$ for $i \neq j$.
The dual model has an $USp(2(\sum_{i=1}^{k} l_i n_i - M)$ gauge group, a similar structure in terms of charged  matter fields and in additions there are further singlets and superpotential interactions. We will be more precise on the structure of the duality below in the body of the paper.

The goal of the analysis consists of giving a physical explanation of this duality finding a sequence of operation that lead to the magnetic phase starting from the electric one.
As anticipated above the possible operations in the sequence are Berkooz's deconfinement and Intriligator-Pouliot dualities \cite{Intriligator:1995ne}, and in addition, when necessary, integrating out massive deformations if they are present or trigger the Higgsing enforced by the (possible) presence in some phase of F-terms giving non trivial VEV to the charged matter fields.

In this way we will be able to derive the duality independently from the integral identity, giving an independent and physical argument in favor of its existence.
As a bonus the sequence discussed above can be entirely applied to the superconformal index, giving origin to a mathematically independent proof of the identity of \cite{rains2012elliptic}.

\section{The duality}
\label{secduality}

In this section we review the identity conjectured in \cite{rains2012elliptic}, then proven in \cite{vandebult2009elliptic} and then the IR duality conjectured in 
\cite{Spiridonov:2009za} that follows from this identity.
The identity relates two hypergeometric elliptic integrals.
Such type of integrals correspond to the supersymmetric indices of the theories in question, as originally shown in \cite{Dolan:2008qi} (where the original indices of \cite{Kinney:2005ej,Romelsberger:2005eg}   were reformulated in the language of elliptic Gamma functions).
In the rest of this paper we will use then the conventions of \cite{Dolan:2008qi,Spiridonov:2009za} for the index and refer the reader to such papers for details.

The identity of \cite{rains2012elliptic} relates
    \begin{equation}
    \begin{split}
        I_{ele} = & \frac{(p,p)_{\infty}^M (q,q)_{\infty}^M}{2^M M!} \Gamma (t;p,q)^{M} \int_{\mathbb{T}^M} \prod_{1 \leq i < j \leq M} \frac{\Gamma(t z_i^{\pm 1} z_j^{\pm 1}; p,q)}{\Gamma (z_i^{\pm 1} z_j^{\pm 1};p,q) \prod_{j=1}^M \Gamma(z_j^{\pm 1};p,q)} \\
        & \prod_{j=1}^M \frac{\prod_{k=1}^4 \Gamma(t t_k^{-1} z_j^{\pm 1}; p,q) \prod_{r=1}^K \prod_{i=1}^{l_r} \Gamma (s_{r,i} z_j^{\pm 1};p,q)}{\prod_{r=1}^K \prod_{i=1}^{l_r} \Gamma(t^{n_r} s_{r,i} z_j^{\pm 1};p,q)} \frac{d z_j}{2 \pi i z_j},
    \end{split}
    \label{Eq:I_Ereview}
    \end{equation}
and
\begin{equation}
            \begin{split}
                & I_{mag} = \frac{(p,p)_{\infty}^N (q,q)_{\infty}^N}{2^N N!} \Gamma (t;p,q)^{N} \prod_{i=0}^{M-N-1} \prod_{1 \leq k < r \leq 4} \Gamma(t^{i+2} t_k^{-1} t_r^{-1};p,q)  \\
                & \prod_{r=1}^{4} \prod_{m=1}^{K} \prod_{i=1}^{l_m} \prod_{k_m=0}^{n_m -1} \frac{\Gamma(t^{k_m +1} t_r^{-1} s_{m,i};p,q)}{\Gamma(t^{k_m} t_r s_{m,i};p,q)} \int_{\mathbb{T}^N} \prod_{1 \leq i < j \leq N} \frac{\Gamma(t z_i^{\pm 1} z_j^{\pm 1}; p,q)}{\Gamma (z_i^{\pm 1} z_j^{\pm 1};p,q) \prod_{j=1}^N \Gamma(z_j^{\pm 2};p,q)} \\ 
                & \prod_{j=1}^N \prod_{k=1}^4 \Gamma (t_k z_j^{\pm 1};p,q) \frac{\prod_{r=1}^K \prod_{i=1}^{l_r} \Gamma (s_{r,i} z_j^{\pm 1};p,q)}{\prod_{r=1}^K \prod_{i=1}^{l_r} \Gamma (t^{n_r} s_{r,i} z_j^{\pm 1};p,q)} \prod_{j=1}^N \frac{\text{d}z_j}{2 \pi z_j}, \\
            \end{split}
            \label{Eq:I_Mreview}
        \end{equation}
where $j=0,\dots,M-N-1$, $k_i=0,\dots,n_i-1$ for any $i=1,\dots,K$, $M+N = \sum_{i=1}^{k} l_i n_i$
(with $M\geq N$) and $\prod_{k=1}^4 t_k = t^{M-N+2}$.

We referred to the integrals in (\ref{Eq:I_Ereview}) and in 
(\ref{Eq:I_Mreview}) as  $I_{ele}$ and $I_{mag}$ because we interpret them as the 
supersymmetric indices of an 
“electric” and a dual “magnetic”  theory, in analogy with the case of ordinary Seiberg  duality.
It is indeed possible to obtain the integrals (\ref{Eq:I_Ereview}) and (\ref{Eq:I_Mreview}) as supersymmetric indices of two distinct 4d $\mathcal{N}=1$ 
supersymmetric gauge theories, the ones the we will denote as the electric and the magnetic phase. 

The electric phase consists in $USp(2M)$ SQCD with a $SU(4) \times USp(2l_1) \times ... \times USp(2l_K)$ flavour symmetry, under which the charged fields are respectively $W, Q_1,..., Q_K$. In addition there is a rank-2 antisymmetric tensor $A$. This is a non-anomalous asymptotically free theory for $M > \sum_{i=1}^K l_i/2-1$ with the superpotential deformation 
    \begin{equation}
    \label{Weleaq}
        W_{\text{el}} = \sum_{i=1}^{K} A^{n_i} Q_i^2
    \end{equation}
turned on. This deformation breaks the would-be $SU(4+\sum_i 2l_i)$ flavour symmetry groups into $SU(4) \times \prod_i USp(2l_i)$. The representations of the fields and their charges under the gauge and flavour groups are summarized in the following table, where $M+N= \sum_{i=1}^K l_i n_i$ and $n_1 \neq n_2 \neq ... \neq n_K$.
 \begin{table}[h]
        \centering
            \begin{tabular}{c|c c c c c c c}
                & $USp(2M)$ & $SU(4)$ & $USp(2l_1)$ & $...$ & $USp(2l_K)$ & $U(1)$ & $U(1)_R$ \\
                \hline
                $W_1$ & $f$ & $\bar{f}$ & $1$ & $...$ & $1$ & $- \frac{N-M-2}{4}$ & $0$ \\
                $Q_1$ & $f$ & $1$ & $f$ & $...$ & $1$ & $- \frac{n_1}{2}$ & $1$ \\
                $...$ &     &     &     &       &     &                   &     \\
                $Q_K$ & $f$ & $1$ & $1$ & $...$ & $f$ & $- \frac{n_K}{2}$ & $1$ \\
                $A$ & $T_A$ & $1$ & $1$ & $...$ & $1$ & $1$               & $0$ \\
                \end{tabular}
                \caption{Field content of the electric theory.}
                \label{Tab:Field_content_el_dual}
    \end{table}

The magnetic phase consists in $USp(2N = 2(\sum_{i=1}^{K} l_i n_i-M))$ SQCD with a $SU(4) \times USp(2l_1) \times ... \times USp(2l_K)$ flavour symmetry, under which the charged fields are respectively $w, q_1,..., q_K$. In addition there is a rank-2 antisymmetric tensor $a$. This is a non-anomalous asymptotically free theory for $N > \sum_{i=1}^K l_i/2-1$. This is anyway not a strict requirement, because we are not focusing here to the 
existence of a duality in the conformal window, but, motivated by the case of ordinary Seiberg duality, we allow also for the possibility of an infrared duality 
between an UV free theory and an IR free one.

We can also write a generic superpotential compatible with the global symmetry structure and compatible with the integral identity shown above. Such 
superpotential for generic values of the 
parameters is  
    \begin{equation}
    \label{pippo}
            \begin{split}
                W_{\text{mag}} = & \sum_{i=1}^{K} a^{n_i} q_i^2 + \sum_{\{i\}} N_{n_i - 2} w_1^2 q_i^2 + \sum_{i=1}^{K} \sum_{\{k_i+l_i \geq n_i-1\}} N_{M-N-1 + n_i -1-k_i-l_i} M_{i,k_i} M_{i, l_i}  \\
                & + \sum_{i=1}^K \sum_{\{k_i + j_i \leq n_i+2\}} M_{i,k_i} M_{i,j_i} w_1^2 a^{n_i -k_i-j_i -2} + \sum_{i=1}^K \sum_{k_i=0}^{n_i-1} M_{i,k_i} w_1 q_i a^{n_i-1-k_i} \\
                & + \sum_{i=1}^K \sum_{\{j \leq n_i-2\}} N_j w_1^2 q_1^2 A^{n_i-2-j} 
            \end{split}
    \end{equation}
 where $N_j$, $j=0,..., M-N-1$, and $M_{i,k_i}$, $k_i =0,...,n_i-1$, are the gauge singlets of the theory, (observe that the $N_j$ fields disappear in the case of $N=M$) and $\sum_{\{i\}}$, $\sum_{\{k_i\}}$, $\sum_{\{j\}}$ are sums over the allowed $i$, $k_i$ and $j$. Besides the terms with $a^{n_i}$ exist only for $N > n_i$ and we can note that all the fields $N_j$ partecipate in the superpotential once we impose the asymptotic freedom of the electric theory. 
 Actually we will flip the singlets of the electric theory and in this way we will not construct from our procedure a dual superpotential in the form of (\ref{pippo}).
 Indeed we will see that the dual superpotential will 
 always vanish after integrating out the singlets and the flippers in the final step of the derivation.
 
 The field content of the dual phase is summarized in the following table
    \begin{table}[h]
        \centering
            \begin{tabular}{c| c c c c c c c}
                & $USp(2N)$ & $SU(4)$ & $USp(2l_1)$ & $...$ & $USp(2l_K)$ & $U(1)$ & $U(1)_R$ \\
                \hline
                $w_1$ & $f$ & $f$ & $1$ & $...$ & $1$ & $ \frac{N-M+2}{4}$ & $0$ \\
                $q_1$ & $f$ & $1$ & $f$ & $...$ & $1$ & $- \frac{n_1}{2}$  & $1$ \\
                $...$ &     &     &      &      &     &                          \\
                $q_K$ & $f$ & $1$ & $1$ & $...$ & $f$ & $- \frac{n_K}{2}$  & $1$ \\
                $N_j$ & $1$ & $\Bar{T}_A$ & $1$ & $...$ & $1$ & $j - \frac{N-M-2}{2}$ & $0$ \\
                $M_{1,k_1}$ & $1$ & $\Bar{f}$ & $f$  & $...$ & $1$ & $k_1\!-\! \frac{N\!-\!M\!-\!2}{4} \!-\! \frac{n_1}{2} $ & $1$ \\
                $...$ &     &     &      &      &     &                          \\
                $M_{K,k_K}$ & $1$ & $\Bar{f}$ & $1$ & $...$ & $f$ & $k_K\!-\! \frac{N\!-\!M\!-\!2}{4} \!-\! \frac{n_K}{2} $ & $1$ \\
                $Y$ & $T_A$ & $1$ & $1$ & $...$ & $1$ & $1$ & $0$ \\    
                \end{tabular}
                \caption{Field content of the magnetic theory.}
                \label{Tab:Field_content_mag_dual}
    \end{table}

This duality has also an interesting limiting (confining) case, given by the choice $N=0$. In the sections below we will provide a derivation of this duality starting in section \ref{sec:confiningcase} by studying the confining case  and then we will move to the more general case in section \ref{sec:dualcase}. Indeed, even if simpler, the essential logic of our derivation is almost completely visible in the confining case, and the few technical differences that appear in the case $N \neq 0$ will be discussed in section \ref{sec:dualcase}.

    \newpage

\section{The Confining limit of the duality}
\label{sec:confiningcase}

In this section we derive the confining case of the  duality from a physical approach, the proof will be recursive. The procedure consists in deconfining a rank-2 antisymmetric tensor with an auxiliary symplectic gauge group and, by sequentially applying infrared dualities, we bring back the dual to the same confining theory with lower rank. In this step we will use the lower rank confining case, which encodes the Higgs mechanism that completely breaks the auxiliary gauge group. This leads to an s-confining gauge theory (namely $USp(2N)$ with $2N+4$ fundamentals). After confining this theory we eventually find the expected WZ model, describing the magnetic phase of the  duality. The first steps of the proof consist in treating some limiting case which shows the presence of the Higgs mechanism in action and help in the construction of the proof to the general rank case. In order to avoid the proliferation of various term in Higgsing we will flip all the would-be mesons of the dual theory. 

The analysis is supported at each step by the relative (integral) identities matching the 4d supersymmetric index. On one hand this corroborates the validity of the results and on the other hand it provides an alternative derivation of the integral identity of \cite{rains2012elliptic,vandebult2009elliptic}.

Let us start the analysis by discussing the gauge theory that can be read from the  duality. It consists in $USp(2M)$ SQCD with a $SU(4) \times USp(2l_1) \times ... \times USp(2l_K)$ flavour symmetry, under which the charged fields are respectively $W, Q_1,..., Q_K$. In addition there is a rank-2 antisymmetric tensor $A$. This theory becomes confining if the superpotential deformation  (\ref{Weleaq})
is turned on.  The representations of the fields and their charges under the gauge and flavour groups are summarized in the following table, where $M= \sum_{i=1}^K l_i n_i$ and $n_1 \neq n_2 \neq ... \neq n_K$.
    \begin{table}[h]
        \centering
            \begin{tabular}{c|c c c c c c c}
                & $USp(2M)$ & $SU(4)$ & $USp(2l_1)$ & $...$ & $USp(2l_K)$ & $U(1)$ & $U(1)_R$ \\
                \hline
                $W$ & $f$ & $\bar{f}$ & $1$ & $...$ & $1$ & $- \frac{M-2}{4}$ & $0$ \\
                $Q_1$ & $f$ & $1$ & $f$ & $...$ & $1$ & $- \frac{n_1}{2}$ & $1$ \\
                $...$ &     &     &     &       &     &                   &     \\
                $Q_K$ & $f$ & $1$ & $1$ & $...$ & $f$ & $- \frac{n_K}{2}$ & $1$ \\
                $A$ & $T_A$ & $1$ & $1$ & $...$ & $1$ & $1$               & $0$ \\
            \end{tabular}
    \end{table}

The integral identity $I_E = I_M$ between these two theories has been explicitly calculated and the indices are the following
    \begin{equation}
    \begin{split}
        I_E = & \frac{(p,p)_{\infty}^M (q,q)_{\infty}^M}{2^M M!} \Gamma (t;p,q)^{M} \int_{\mathbb{T}^M} \prod_{1 \leq i < j \leq M} \frac{\Gamma(t z_i^{\pm 1} z_j^{\pm 1}; p,q)}{\Gamma (z_i^{\pm 1} z_j^{\pm 1};p,q) \prod_{j=1}^M \Gamma(z_j^{\pm 1};p,q)} \\
        & \prod_{j=1}^M \frac{\prod_{k=1}^4 \Gamma(t t_k^{-1} z_j^{\pm 1}; p,q) \prod_{r=1}^K \prod_{i=1}^{l_r} \Gamma (s_{r,i} z_j^{\pm 1};p,q)}{\prod_{r=1}^K \prod_{i=1}^{l_r} \Gamma(t^{n_r} s_{r,i} z_j^{\pm 1};p,q)} \frac{d z_j}{2 \pi i z_j},
    \end{split}
    \label{Eq:I_E}
    \end{equation}
    
    \begin{equation}
        I_M = \prod_{i=0}^{M-1} \prod_{1 \leq k < r \leq 4} \Gamma(t^{i+2} t_k^{-1} t_r^{-1};p,q) \prod_{r=1}^{4} \prod_{m=1}^{K} \prod_{i=1}^{l_m} \prod_{k_m=0}^{n_m -1} \frac{\Gamma(t^{k_m +1} t_r^{-1} s_{m,i};p,q)}{\Gamma(t^{k_m} t_r s_{m,i};p,q)},
    \label{Eq:I_M}
    \end{equation}
with the constraint $\prod_{k=1}^4 t_k = t^{M+2}$. From the magnetic index we read following magnetic theory, where $k_i = 0,..., n_i-1$ for $i=1,...,K$ and $j=0,..., M-1$.
    \begin{table}[ht]
        \centering
        \begin{tabular}{c| c c c c c c}
            & $SU(4)$ & $USp(2l_1)$ & $...$ & $USp(2l_K)$ & $U(1)$ & $U(1)_R$ \\
            \hline
            $W^2 A^j$       & $\Bar{T}_A$ & $1$ & $...$ & $1$ & $j - \frac{M-2}{2}$                    & $0$ \\
            $W Q_1 A^{k_1}$ & $\Bar{f}$   & $f$ & $...$ & $1$ & $- \frac{M-2}{4} - \frac{n_1}{2} +k_1$ & $1$ \\
            $...$             &             &     &       &     &                                        &     \\
            $W Q_K A^{k_K}$ & $\Bar{f}$   & $1$ & $...$ & $f$ & $- \frac{M-2}{4} - \frac{n_K}{2} +k_K$ & $1$ \\
        \end{tabular}
    \end{table}

\subsection{Field theory approach}

We begin our analysis with a field theory approach, in order to implement the recursive program we need to discuss some limiting case, that will be used as base steps of the recursion.

\paragraph{$ \bullet \quad l_1 = n_1 = 1$ case:} 
    This case is trivial, since the anti-symmetric tensor $A$ is a singlet and we have an $USp(2)$ gauge group with $6$ fundamentals, which is the confining case of Intrilligator-Pouliot duality. 

\begin{figure}[ht]
    \centering
    \begin{tikzpicture}[shorten >=1pt,shorten <=1pt,node distance=4cm,on grid,auto]
        \node[rectangle, draw=red!50, fill=red!20, inner sep=2pt, minimum size=5mm] (1) at (-3,0) {$4$};
        \node[circle, draw=blue!50, fill=blue!20, inner sep=2pt, minimum size=5mm] (2) at (0,0) {$4$};
        \node[rectangle, draw=blue!50, fill=blue!20, inner sep=2pt, minimum size=5mm] (3) at (3,0) {$4$};
        \draw
        (1) edge[loop above] node {$M_1^{'}$} (1)
        (1) edge[loop left] node {$M^{'}_2$} (1)
        (2) edge[loop above] node {$A$} (2)
        (3) edge[loop above] node {$M_1$} (3)
        (3) edge[loop right] node {$M_2$} (3);
        \draw[->]
        (1) edge[auto shift] node[above] {$w$} (2);
        \draw (2) -- node {$q$} (3);
        \node at (0,-1) {$W = M_1 + M_1 q^2 + M_1^{'} w^2 + M_2 q^2 A + M_2^{'} w^2 A $};
    \end{tikzpicture}
    \caption{Quiver representation of the  electric confining theory $l_1 =2$ and $n_1=1$ after having turned on the superpotential deformation in the dual theory. Gauge groups are represented as circles while flavor
    nodes are represented with squares. Symplectic groups are depicted in blue and unitary groups are depicted in red. In-going and outgoing arrows are respectively fundamental and anti-fundamentals fields, loops are antisymmetric tensor fields.}
    \label{Q:SWV_el_Higgsing_l1=2_n1=1}
    \end{figure}
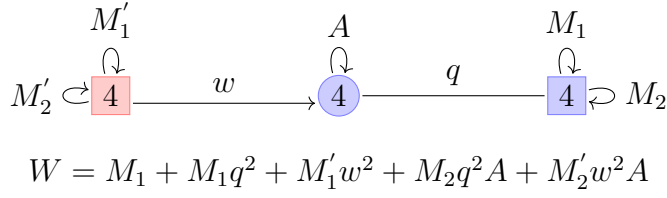

\paragraph{$ \bullet \quad l_1 = 2$ and $n_1 = 1$ case:}
    The theory is a $USp(4)$ gauge theory with flavor group $SU(4) \times USp(4)$. Without adding the deformation $W_{el} = A Q_1^2$ the theory has the following global symmetry $SU(4)_1 \times SU(4)_2 \times U(1)_a \times U(1)_b \times U(1)_{\Tilde{R}} $. We consider one of the $72$ dual phases of the latter theory, precisely the one reported in Table $(2.9)$ of \cite{Razamat:2017hda}, and we turn on the interaction, which breaks the global symmetry to $USp(4) \times SU(4) \times U(1) \times U(1)_R$, mixing the Abelian symmetries in the following way
        \begin{equation}
        \begin{split}
            & U(1) = U(1)_a - \frac{1}{4} U(1)_b, \\
            & U(1)_R = U(1)_{\Tilde{R}} + \frac{1}{2} U(1)_b.
        \end{split}
        \label{Eq:U(1)_breaking}
        \end{equation}
    We depicted the model in Figure \ref{Q:SWV_el_Higgsing_l1=2_n1=1} in terms of a quiver gauge theory.

    The dictionary between these new fields and the previous ones is 
        \begin{equation}
            M_{2-l} \longleftrightarrow Q_1^2 A^l, \hspace{1 cm} M_{2-l}^{'} \longleftrightarrow W^2 A^l, \hspace{1 cm} A \longleftrightarrow A,
        \end{equation}
    where $l= 0,1$ and the electric mesons $WQ_1$ are mapped in the magnetic mesons $wq$.
    We see from the superpotential reported in Figure \ref{Q:SWV_el_Higgsing_l1=2_n1=1} that $M_1$ is linear and its F-term gives vev to $q^2$. In fact by taking $q$ diagonal in the color-flavor space 
        \begin{equation}
            q_r^I = \begin{pmatrix}
            \lambda_{1} & \\
            & \lambda_{2}
            \end{pmatrix} \otimes i \sigma_2,
        \end{equation}
    the equation of motion of $M_1$ constrain $q$ to solve
        \begin{equation}
            qq = q_a^I q^J_b J_{IJ}^{\text{gauge}} = - J^{\text{flav}}_{ab} = -(\mathbb{I}_{2} \otimes i \sigma_2)_{ab},
        \end{equation}
    which implies
        \begin{equation}
            q = \mathbb{I}_{2} \otimes i \sigma_2.
        \end{equation}
    This vev breaks the gauge-flavor symmetry $USp(4) \times USp(4)$ to the flavor diagonal subgroup $USp(4)$. The superpotential after this breaking is 
        \begin{equation}
            W = M_1^{'} w^2 + M_2 A + M_2^{'} w^2 A,
        \end{equation}
    where now $A$ is the anti-symmetric field of the $USp(4)$ flavor group and $w \longleftrightarrow W Q_1$. From this potential we see that both $M_2$ and $A$ are massive and can be integrated out. This conclude the analysis of this case.

\paragraph{$ \bullet \quad l_1 = 2 $ and $ n_1 = n $ case:}
    The procedure is the same of the previous case and after adding the deformation $W_{el} = A^n Q_1^2$ the breaking pattern of symmetries is the one in eq. \eqref{Eq:U(1)_breaking}. We depicted this model in Figure \ref{Q:SWV_el_Higgsing_l1=2_n1=n}.
    
    \begin{figure}[ht]
    \centering
    \begin{tikzpicture}[shorten >=1pt,shorten <=1pt,node distance=4cm,on grid,auto]
        \node[rectangle, draw=red!50, fill=red!20, inner sep=2pt, minimum size=5mm] (1) at (-3,0) {$4$};
        \node[circle, draw=blue!50, fill=blue!20, inner sep=2pt, minimum size=5mm] (2) at (0,0) {$4N$};
        \node[rectangle, draw=blue!50, fill=blue!20, inner sep=2pt, minimum size=5mm] (3) at (3,0) {$4$};
        \node at (-3,1.1) {$\vdots$};
        \node at (3,1.1) {$\vdots$};
        \draw
        (1) edge[loop above] node[right] {$M_1^{'}$} (1)
        (1) edge[loop above, min distance=15mm] node[right] {$M^{'}_{2n}$} (1)
        (2) edge[loop above] node {$A$} (2)
        (3) edge[loop above] node[right] {$M_1$} (3)
        (3) edge[loop above, min distance=15mm] node[right] {$M_{2n}$} (3);
        \draw[->]
        (1) edge[auto shift] node {$w$} (2);
        \draw (2) -- node {$q$} (3);
        \node at (0,-1) {$W = M_n + \sum_{l=1}^{2n} M_l q^2 A^{l-1} + \sum_{l=1}^{2n} M_l^{'} w^2 A^{l-1} $};
    \end{tikzpicture}
    \caption{Quiver representation of the  electric confining theory $l_1 =2$ and $n_1=n$ after having turned on the superpotential deformation $A^n Q^2$ in the dual theory. }
    \label{Q:SWV_el_Higgsing_l1=2_n1=n}
    \end{figure}
    The dictionary between these new fields and the electric ones is the following 
        \begin{equation}
            M_{2n-l} \longleftrightarrow Q^2 A^l, \hspace{1 cm} M_{2n-l}^{'} \longleftrightarrow W^2 A^l, \hspace{1 cm} A \longleftrightarrow A,
        \label{Eq:Duality_Mapping_72_Duals}
        \end{equation}
    where $l=0,...,2n-1$ and the electric mesons are mapped into the magnetic mesons. We can note from the superpotential in Figure \ref{Q:SWV_el_Higgsing_l1=2_n1=n} that $M_n$ is linear and its F-term gives a vev to $q^2 A^{n-1}$. This vev breaks completely the $USp(4N)$ gauge group and the $USp(4)$ flavour group to the diagonal $USp(4)$ flavour subgroup. Expanding around this vev we can see that all the $M_l$ fields and the antisymmetric tensor $A$ become massive and the original $w$ fields decompose into the $W Q_1 A^{k_1}$ fields. 

\begin{figure}[ht]
    \centering
    \begin{tikzpicture}[shorten >=1pt,shorten <=1pt,node distance=4cm,on grid,auto]
        \node[rectangle, draw=red!50, fill=red!20, inner sep=2pt, minimum size=5mm] (1) at (-3,0) {$3$};
        \node[rectangle, draw=red!50, fill=red!20, inner sep=2pt, minimum size=5mm] (4) at (-3,2) {$1$};
        \node[circle, draw=blue!50, fill=blue!20, inner sep=2pt, minimum size=5mm] (2) at (0,0) {$6$};
        \node[rectangle, draw=blue!50, fill=blue!20, inner sep=2pt, minimum size=5mm] (3) at (3,0) {$6$};
        \node[circle, draw=blue!50, fill=blue!20, inner sep=2pt, minimum size=5mm] (5) at (0,2) {$4$};
        \node[rectangle, draw=red!50, fill=red!20, inner sep=2pt, minimum size=5mm] (6) at (3,2) {$1$};
        \draw[->]
        (1) edge[auto shift] node {$W_{2}$} (2)
        (1) edge[auto shift] node[left] {$\beta_j$} (4);
        \draw (1) edge[loop left] node {$\alpha_j$} (1);
        \draw (2) -- node [left]{$B$} (5);
        \draw (2) -- node [right]{$D$} (6);
        \draw (2) -- node [above]{$Q_1$} (3);
        \draw (5) -- node [below]{$F$} (4);
        \draw (6) -- node [below]{$C$} (5);
        \draw  -- (4) edge[bend left=45] node [above]{$E$}(6);
        \node at (0,-1.5) {$W = (BQ_{1})^2 + BCD + CEF + \theta \text{Tr}(B^2) + \sum_{j=0}^1 (\alpha_j (W_2 B (B^2)^j F) + \alpha_2 W_2 D)$};
        \node at (0,-2.5) {$+ \sum_{j=0}^2 \beta_j (W_2^2 (B^2)^j))$};
    \end{tikzpicture}
    \caption{Quiver representation of the  electric confining theory $l_1 =3$ and $n_1=1$ after the deconfinement of the $USp(6)$ rank-2 anti-symmetric tensor $A$. The $\alpha_j$ and $\beta_j$ in the Figure are the ones in the superpotential.}
    \label{Q:SWV_el_deconfined_l1=3_n1=1}
    \end{figure}
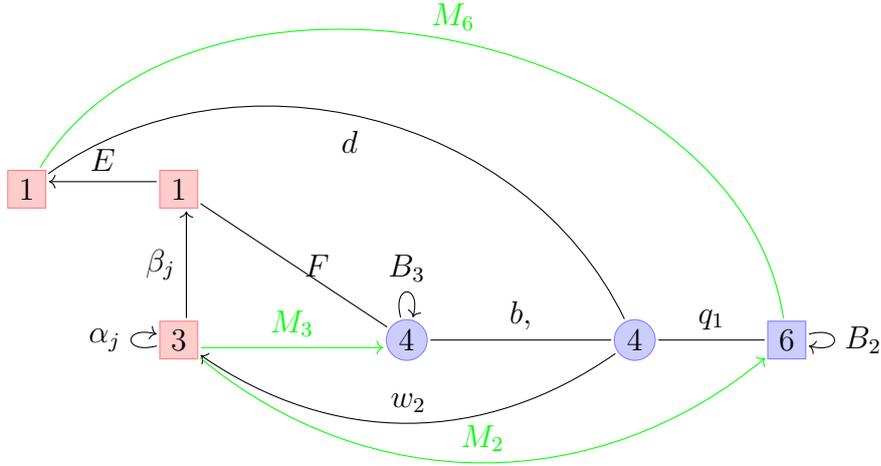

\paragraph{$\bullet \quad l_1=3$ and $n_1=1$ case:}
    In this case the theory is a $USp(6)$ gauge theory with $SU(4) \times USp(4)$ flavor symmetry and the deformation of the electric theory is $W_{el} = A Q_1^2$. We begin by breaking the $SU(4)$ flavour group to $SU(3) \times U(1)$ and by flipping all the would-be mesons of the type $W^2 A^j$, with $j=0,...,M-1$. We represent this new model in Figure \eqref{Q:SWV_el_flav_brkn_l1=3_n1=1}.
    \begin{figure}[h]
    \centering
    \begin{tikzpicture}[shorten >=1pt,shorten <=1pt,node distance=4cm,on grid,auto]
        \node[rectangle, draw=red!50, fill=red!20, inner sep=2pt, minimum size=5mm] (1) at (-3,-1) {$3$};
        \node[rectangle, draw=red!50, fill=red!20, inner sep=2pt, minimum size=5mm] (4) at (-3,1) {$1$};
        \node[circle, draw=blue!50, fill=blue!20, inner sep=2pt, minimum size=5mm] (2) at (0,0) {$6$};
        \node[rectangle, draw=blue!50, fill=blue!20, inner sep=2pt, minimum size=5mm] (3) at (3,0) {$6$};
        \draw
        (2) edge[loop above] node {$A$} (2)
        (1) edge[loop left] node {$\alpha_j$} (1);
        \draw[->]
        (1) edge[auto shift] node {$W_{2}$} (2)
        (4) edge[auto shift] node[above] {$W_{1}$} (2)
        (1) edge[auto shift] node[left] {$\beta_j$} (4);
        \draw (2) -- node [above]{$Q_1$} (3);
        \node at (0,-2) {$W = AQ_{1}^2 + \sum_{j=0}^2 (\alpha_j (W_1 W_2 A^j) + \beta_j (W_2^2 A^j)) $};
    \end{tikzpicture}
    \caption{Quiver representation of the  electric confining theory $l_1 =3$ and $n_1=1$ after having broken the $SU(4)$ flavour group to $SU(3) \times U(1)$. The fields $\alpha_k$ and $\beta_k$ are flipper of the would-be mesons $W^2 A^j$. In the Figure $j=0,1,2$.}
    \label{Q:SWV_el_flav_brkn_l1=3_n1=1}
    \end{figure}
    We now deconfine the anti-symmetric field $A$ using the broken $U(1)$ flavor group. We depicted this step in Figure \ref{Q:SWV_el_deconfined_l1=3_n1=1}.
    The anti-symmetric tensor $A$ is recovered by confining the $USp(4)$ gauge node in terms of the field $B$, i.e. $A \sim B^2$, where the contraction is done on the $USp(4)$ indices.

    The next step consists of Intrilligator-Pouliot duality on $USp(6)$, this gauge group is dual to a $USp(4)$ gauge theory, which is reported in Figure \ref{Q:SWV_el_dualized_l1=3_n1=1}.
    \begin{figure}[h]
    \centering
    \begin{tikzpicture}[shorten >=1pt,shorten <=1pt,node distance=4cm,on grid,auto]
        \node[rectangle, draw=red!50, fill=red!20, inner sep=2pt, minimum size=5mm] (1) at (-3,0) {$3$};
        \node[rectangle, draw=red!50, fill=red!20, inner sep=2pt, minimum size=5mm] (4) at (-3,2) {$1$};
        \node[circle, draw=blue!50, fill=blue!20, inner sep=2pt, minimum size=5mm] (2) at (3,0) {$4$};
        \node[rectangle, draw=blue!50, fill=blue!20, inner sep=2pt, minimum size=5mm] (3) at (5,0) {$6$};
        \node[circle, draw=blue!50, fill=blue!20, inner sep=2pt, minimum size=5mm] (5) at (0,0) {$4$};
        \node[rectangle, draw=red!50, fill=red!20, inner sep=2pt, minimum size=5mm] (6) at (-5,2) {$1$};
        \draw
        (1) edge[loop left] node {$\alpha_j$} (1)
        (5) edge[loop above] node {$B_3$} (5)
        (3) edge[loop right] node {$B_2$} (3);
        \draw [->] (2) edge[bend left = 35] node[above] {$w_{2}$} (1);
        \draw [->] (1) edge[bend right = 40, color = green] node {$M_2$} (3);
        \draw [->] (1) edge[auto shift, color = green] node [above] {$M_3$} (5);
        \draw [->] (1) edge[auto shift] node [left] {$\beta_j$} (4);
        \draw (2) -- node [above]{$b,$} (5);
        \draw -- (2) edge[bend right= 50] node {$d$} (6);
        \draw -- (3) edge[bend right= 70, color = green] node [above]{$M_6$} (6);
        \draw (2) -- node [above]{$q_1$} (3);
        \draw (5) -- node [right]{$F$} (4);
        \draw  -- (4) edge[auto shift] node [above]{$E$}(6);
        \node at (0,-2.5) {$W = B_3 b^2 + M_3 w_2 b + (bq_1)^2 + dbEF + \theta \text{Tr}(B_3) + M_1 w_2 d + M_2 w_2 q_1 + M_6 d q_1 $};
        \node at (0,-3.5) {$ + B_2 q_1^2 + \alpha_0 M_3 F + \alpha_1 M_3 B_3 F  + \beta_1 M_3^2 + \beta_2 M_3^2 B_3$};
    \end{tikzpicture}
    \caption{Quiver representation of the  electric confining theory $l_1 =3$ and $n_1=1$ after Intriligator-Pouliot duality on $USp(6)$. The rank of the dual gauge group has been lowered to $4$ and new mesonic degrees of freedom appear. We highlighted the non-massive mesons in green. In the Figure the $\alpha_j$ and $\beta_j$ are the ones reported in the superpotential.}
    \label{Q:SWV_el_dualized_l1=3_n1=1}
    \end{figure}
    The dictionary between these new fields and the previous ones is the following:
        \begin{equation}
            \begin{split}
                & M_1 \longleftrightarrow W_2 D, \hspace{1 cm} M_2 \longleftrightarrow W_2 Q_1, \hspace{1 cm} M_3 \longleftrightarrow W_2 B, \hspace{1 cm} M_4 \longleftrightarrow B D, \\
                & M_5 \longleftrightarrow B Q_1, \hspace{1 cm} M_6 \longleftrightarrow D Q_1, \hspace{1 cm} B_1 \longleftrightarrow (W_2)^2, \hspace{1 cm} B_2 \longleftrightarrow Q_1^{2}, \\
                & B_3 \longleftrightarrow B^{2}, \\
            \end{split} 
        \end{equation}
    where the fields $C, M_4,  M_5, M_1, B_1, \alpha_2, \beta_0$ are massive and have been resolved in favor of their equations of motion.

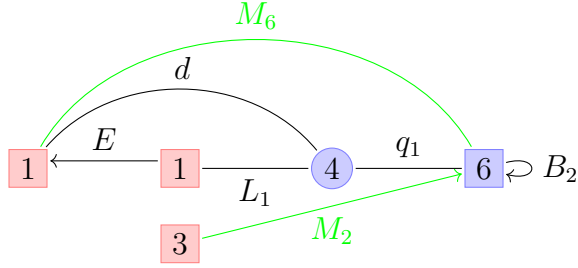
    \begin{figure}[h]
    \centering
    \begin{tikzpicture}[shorten >=1pt,shorten <=1pt,node distance=4cm,on grid,auto]
        \node[rectangle, draw=red!50, fill=red!20, inner sep=2pt, minimum size=5mm] (1) at (-1,-1) {$3$};
        \node[rectangle, draw=red!50, fill=red!20, inner sep=2pt, minimum size=5mm] (4) at (-1,0) {$1$};
        \node[circle, draw=blue!50, fill=blue!20, inner sep=2pt, minimum size=5mm] (2) at (1,0) {$4$};
        \node[rectangle, draw=blue!50, fill=blue!20, inner sep=2pt, minimum size=5mm] (3) at (3,0) {$6$};
        \node[rectangle, draw=red!50, fill=red!20, inner sep=2pt, minimum size=5mm] (6) at (-3,0) {$1$};
        \draw
        (3) edge[loop right] node {$B_2$} (3);
        \draw [->] (1) edge[color = green] node[below] {$M_2$} (3);
        \draw -- (2) edge[bend right= 50] node [above]{$d$} (6);
        \draw -- (3) edge[bend right= 60, color = green] node [above]{$M_6$} (6);
        \draw (2) -- node [above]{$q_1$} (3);
        \draw (2) -- node [below]{$L_1$} (4);
        \draw  -- (4) edge[auto shift] node [above]{$E$}(6);
        \node at (0,-2) { $W = dL_1E + M_6 d q_1 + B_2 q_1^2$};
    \end{tikzpicture}
    \caption{Quiver representation of the  electric confining theory $l_1 =3$ and $n_1=1$ after the recursive step and having integrated out the massive fields.}
    \label{Q:SWV_el_confined_l1=3_n1=1}
    \end{figure}

    Now, we note that the auxiliary $USp(4)$ gauge node present the same structure of the  theory we started with, but with $l_1^{'}=2$ and $n_1^{'}=1$. Hence we traced back the theory to a previous case. Then, by applying the result of the last paragraph and by integrating out the massive fields, we get the theory depicted in Figure \ref{Q:SWV_el_confined_l1=3_n1=1}.
    The $USp(4)$ gauge node with $8$ fundamentals is the confining case of Intriligator-Pouliot duality. The fields $E, M_6, B_2$ get masses and are integrated out. The only surviving fields are $M_2$ and the meson $L_1 q_1$, which are the not-flipped mesons of the magnetic theory.

\paragraph{$\bullet \quad l_1=l$ and $n_1=1$ case:}
    The procedure is the very same of the previous paragraph. We deconfine the two-index tensor $A$ thanks to a confining auxiliary gauge group $USp(2l-2)$ and we flip all the would-be mesons of the type $A^j W^2$, $j=0,...,M-1$. Then we dualize the $USp(2l)$ gauge group and we obtain a sub-quiver in which we recognize the recursive step with $l_1^{'} = l-1$ and $n_1^{'} =1$. Then we use the confined description of this theory and  integrate out the massive fields. This leads to a confining $USp(2l-2)$ gauge theory with $2l+2$ fundamentals. After the last confinement we obtain the claimed magnetic theory.

\paragraph{$\bullet \quad l_1=l$ and $n_1=n$ case:}
    We are now ready to prove this more general case, which encodes all the recursive steps that are needed to prove all the other cases. The theory is a $USp(2nl)$ gauge theory with $SU(4) \times USp(2l)$ flavor group, and the electric superpotential is deformed to $W_{el} = A^n Q_1^2$. After the step of deconfinement and duality we obtain the theory in Figure \ref{Q:SWV_el_dualized_l1=l_n1=n}.
    \begin{figure}[h]
    \centering
    \begin{tikzpicture}[shorten >=1pt,shorten <=1pt,node distance=4cm,on grid,auto]
        \node[rectangle, draw=red!50, fill=red!20, inner sep=2pt, minimum size=5mm] (1) at (-3,0) {$3$};
        \node[rectangle, draw=red!50, fill=red!20, inner sep=2pt, minimum size=5mm] (4) at (-3,2) {$1$};
        \node[circle, draw=blue!50, fill=blue!20, inner sep=2pt, minimum size=5mm] (2) at (3,0) {$2(l-1)$};
        \node[rectangle, draw=blue!50, fill=blue!20, inner sep=2pt, minimum size=5mm] (3) at (5,0) {$2l$};
        \node[circle, draw=blue!50, fill=blue!20, inner sep=2pt, minimum size=5mm] (5) at (0,0) {$2(nl-1)$};
        \node[rectangle, draw=red!50, fill=red!20, inner sep=2pt, minimum size=5mm] (6) at (-5,2) {$1$};
        \draw
        (1) edge[loop left] node {$\alpha_j$} (1)
        (5) edge[loop above] node {$B_3$} (5)
        (3) edge[loop right] node {$B_2$} (3);
        \draw [->] (2) edge[bend left = 40] node[above] {$w_{2}$} (1);
        \draw [->] (1) edge[bend right = 50, color = green] node {$M_2$} (3);
        \draw [->] (1) edge[auto shift, color = green] node [above] {$M_3$} (5);
        \draw [->] (5) edge[bend left = 45 , color = green] node [above] {$M_5$} (3);
        \draw [->] (1) edge[auto shift] node [left] {$\beta_j$} (4);
        \draw (2) -- node [above]{$b$} (5);
        \draw -- (2) edge[bend right= 50] node {$d$} (6);
        \draw -- (3) edge[bend right= 70, color = green] node [above]{$M_6$} (6);
        \draw (2) -- node [above]{$q_1$} (3);
        \draw (5) -- node [right]{$F$} (4);
        \draw  -- (4) edge[auto shift] node [above]{$E$}(6);
        \node at (0,-2.5) {$W = B_3^{n-1} M_5^2 + B_3 b^2 + dbEF + B_2 q^2 + \theta \text{Tr}(B_3) + M_2 w_2 q + M_3 w_2 b + M_5 b q $};
        \node at (0,-3.5) {$ + M_6 d q + \sum_{j=0}^{nl-2} \alpha_j B_3^j M_3 F + \sum_{j=1}^{nl-1} \beta_j B_3^{j-1} M_3^2$};
    \end{tikzpicture}
    \caption{Quiver representation of the  electric confining theory $l_1 =l$ and $n_1=n$ after the deconfinement and the Intriligator-Pouliot duality on the $USp(2nl)$ gauge group. The rank of the dual gauge group has been lowered to $2(nl-1)$ and new mesonic degrees of freedom appear. We highlighted the non-massive mesons in green. In the Figure the $\alpha_j$ and $\beta_j$ are the ones in the superpotential.}
    \label{Q:SWV_el_dualized_l1=l_n1=n}
    \end{figure}
    The theory has been traced back to the same kind of theory with two $USp(2l_i)$ flavor group, namely $USp(2l) \times USp(2(l-1))$. Hence we have a $USp(2(nl-1)$ gauge group theory with $SU(3) \times U(1) \times USp(2l_1^{'}) \times USp(2l_2^{'})$, with $n_1^{'} = n-1$, $l_1^{'} = l$ and $n_2^{'} = 1$ and $l_2^{'} = l-1$. In order to prove with the recursive algorithm this step, we need to prove the case with $n_1=n$, $l_1=l_1$ and $n_2 =1$, $l_2 = l_2$. With the very same steps we have already shown we can trace back this former case to the theory with $n_1= n_2 =1$ and different $l_1, l_2$. But this theory is a pure rewriting of the case with $n_1 = 1$ and $l_1 = l$ already studied above. In general this type of reasoning can be applied to all the cases with $n_i \neq 0$, $l_i \neq 0$, which are traced back to the case with all $n_i=1$, which can be seen as the case with only one $n_i=1$ and generic $l$, which we already proved. So, we can apply the recursive step and we obtain, after integrating out the massive fields, the theory in Figure \ref{Q:SWV_el_confined_l1=l_n1=n}. The dictionary of the confining duality is:
        \begin{equation}
            M_{1,k_1}^{'} \longleftrightarrow M_3 M_5 A^{k_1}, \hspace{1 cm} M_{1,k_1}^{''} \longleftrightarrow F M_5 A^{k_1}, \hspace{1 cm} M_{2,0}^{''} \longleftrightarrow F b,
        \end{equation}
    and $k_1 = 0,..., n-2$.
    \begin{figure}[h]
    \centering
    \begin{tikzpicture}[shorten >=1pt,shorten <=1pt,node distance=4cm,on grid,auto]
        \node[rectangle, draw=red!50, fill=red!20, inner sep=2pt, minimum size=5mm] (1) at (3,-1) {$3$};
        \node[rectangle, draw=red!50, fill=red!20, inner sep=2pt, minimum size=5mm] (4) at (-1,0) {$1$};
        \node[circle, draw=blue!50, fill=blue!20, inner sep=2pt, minimum size=5mm] (2) at (1,0) {$2(l-1)$};
        \node[rectangle, draw=blue!50, fill=blue!20, inner sep=2pt, minimum size=5mm] (3) at (3,0) {$2l$};
        \node[rectangle, draw=red!50, fill=red!20, inner sep=2pt, minimum size=5mm] (6) at (-3,0) {$1$};
        \draw
        (3) edge[loop right] node {$B_2$} (3);
        \draw [->] (1) edge[auto shift] node[right] {$M_2$} (3);
        \draw [->] (1) edge [color = green] node[left] {$M_{1,k_1}^{'}$} (3);
        \draw [->] (4) edge [bend left = 45, color = green] node[above] {$M_{1,k_1}^{''}$} (3);
        \draw -- (2) edge[bend right= 50] node [above]{$d$} (6);
        \draw -- (3) edge[bend right= 60] node [above]{$M_6$} (6);
        \draw (2) -- node [above]{$q_1$} (3);
        \draw -- (2) edge[color = green] node [below]{$M_{2,0}^{''}$} (4);
        \draw  -- (4) edge[auto shift] node [above]{$E$}(6);
        \node at (0,-2) { $W = dE M_{2,0}^{''} + M_6 d q_1 + B_2 q_1^2$};
    \end{tikzpicture}
    \caption{Quiver representation of the  electric confining theory $l_1 =l$ and $n_1=n$ after the recursive step and having integrated out the massive fields. We highlighted the mesonic deformations in green and $k_1=0,..., n-2$.}
    \label{Q:SWV_el_confined_l1=l_n1=n}
    \end{figure}
    We can notice that now the $USp(2(l-1))$ gauge group has $2l+2$ fundamentals and so it is confining. By using the confinement and by integrating out the massive fields we obtain the claimed magnetic theory, where the remaining mesons have the following identification:
        \begin{equation}
            \begin{split}
            & M_2 \longleftarrow W_2 Q_1, \hspace{1 cm} M_{1,k_1}^{'} \longleftrightarrow W_2 Q_1 A^{k_1 +1}, \hspace{1 cm} M_{2,0}^{''} q_1 \longleftrightarrow W_1 Q_1 A^{n-1},\\
            & M_{1,k_1}^{''} \longleftrightarrow W_1 Q_1 A^{k_1},
            \end{split}
        \end{equation}
    where $k_1 = 0,...,n-2$.

    In all other cases, even when we have different $USp(2l_i)$ flavor groups, the procedure is the same, by deconfining the $2-$index anti-symmetric tensor $A$ and dualizing the original gauge group we get a theory from the same family of the ones we considered but with lower ranks. By applying the recursive step we get a Intriligator-Pouliot confining gauge theory. The interested reader can reproduce these results by following the stepwise procedure that we presented above.


\subsection{Approach via the supersymmetric index}
    In this subsection we reproduce the analysis using the $S^3 \times S^1$ supersymmetric index. We exploit this analysis by recursive steps as the previous field approach. We start by rewriting $I_E$ and $I_M$ in formulas \eqref{Eq:I_E} and \eqref{Eq:I_M} by modifying the $USp(2l_r)$ fugacities as $s_{r,i} \longrightarrow (pq)^{1/2} t^{-n_r/2} \Tilde{s}_{r,i}$. This, with the reflection equation
        \begin{equation}
            \Gamma (z;p,q) \Gamma (pq/z;p,q) =1
        \label{Eq:Reflection_Eq}
        \end{equation}
    gives 
        \begin{equation}
            \begin{split}
                I_E = & \frac{(p,p)_{\infty}^M (q,q)_{\infty}^M}{2^M M!} \Gamma (t;p,q)^{M} \int_{\mathbb{T}^M} \prod_{1 \leq i < j \leq M} \frac{\Gamma(t z_i^{\pm 1} z_j^{\pm 1}; p,q)}{\Gamma (z_i^{\pm 1} z_j^{\pm 1};p,q) \prod_{j=1}^M \Gamma(z_j^{\pm 1};p,q)} \\
                & \prod_{j=1}^M \prod_{k=1}^4 \Gamma(t t_k^{-1} z_j^{\pm 1}; p,q) \prod_{r=1}^K \prod_{i=1}^{l_r} \Gamma ((pq)^{1/2} t^{-n_r/2} \tilde{s}_{r,i}^{\pm 1} z_j^{\pm 1};p,q) \prod_{j=1}^{M} \frac{d z_j}{2 \pi i z_j},
            \end{split}
        \end{equation}
    and
        \begin{equation}
            I_M = \prod_{i=0}^{M-1} \prod_{1 \leq k < r \leq 4} \Gamma(t^{i+2} t_k^{-1} t_r^{-1};p,q) \prod_{r=1}^{4} \prod_{m=1}^{K} \prod_{i=1}^{l_m} \prod_{k_m=0}^{n_m -1} \Gamma((pq)^{1/2} t^{k_m +1 - n_m/2} t_r^{-1} s_{m,i}^{\pm 1};p,q),
        \end{equation}
    again with the balancing condition $\prod_{k=1}^4 t_k = t^{M+2}$.
\paragraph{$\bullet \quad l_1 = 2$ and $n_1 = n$ case:}
    We consider the theory without the deformation $A^n Q^2$, so we can work with one of the $72$ dual phases of \cite{Razamat:2017hda} reported in Figure \ref{Q:SWV_el_Higgsing_l1=2_n1=n}. The index is (in the following we will use the notation $\Gamma (z;p,q) = \Gamma (z)$)
        \begin{equation}
            \begin{split}
                I_E = & \frac{(p,p)_{\infty}^{2n} (q,q)_{\infty}^{2n}}{2^{2n} 2n!} \Gamma(t)^{2n} \prod_{l=1}^{2n} \prod_{1 \leq u < v \leq 4} \Gamma ((pq)^{1/2} v^{2} u^{(2n+1)/2 -l} x_u x_v)  \\
                & \prod_{1 \leq a < b \leq 4} \Gamma ((pq)^{1/2} v^{-2} u^{(2n+1)/2 -l} s_a s_b) \int_{\mathbb{T}^{2n}} \frac{\prod_{1 \leq i < j \leq 2n} \Gamma(u z_i^{\pm 1} z_j^{\pm 1})}{\prod_{i=1}^{2n} \Gamma(z_i^{\pm 2}) \prod_{1 \leq i < j \leq 2n} \Gamma (z_i^{\pm 1} z_j^{\pm 1})} \\
                & \prod_{u=1}^4 \prod_{i=1}^{2n} \Gamma((pq)^{1/4} v^{-1} u^{- \frac{2n-1}{4}} x_u z_i^{\pm 1}) \prod_{a=1}^4 \Gamma ((pq)^{1/4} v u^{-\frac{2n-1}{4}} z_i^{\pm 1} s_a^{-1}) \prod_{i=1}^{2n} \frac{dz_i}{2 \pi i z_i},
            \end{split}
        \end{equation}
    where the dependence on the additional Abelian $U(1)$ symmetry is exploited by the fugacity $v$ and the balancing conditions are $\prod_{a=1}^4 s_a = 1$ and $\prod_{u=1}^4 x_u = 1$. We can note that the following combinations of gamma functions 
        \begin{equation}
            \begin{split}
                & \prod_{j=1}^{n-1} \Gamma(u z_j z_{j+1}^{-1}) \Gamma ((pq)^{1/4} v^{-1} u^{-(2n-1)/4} z_1^{-1} x_1) \Gamma ((pq)^{1/4} v^{-1} u^{-(2n-1)/4} z_n x_4), \\
                & \prod_{j=n+1}^{2n-1} \Gamma(u z_j z_{j+1}^{-1}) \Gamma ((pq)^{1/4} v^{-1} u^{-(2n-1)/4} z_{n+1}^{-1} x_2) \Gamma ((pq)^{1/4} v^{-1} u^{-(2n-1)/4} z_{2n} x_3), \\
            \end{split} 
        \label{Eq:Gamma_Combination}
        \end{equation}
    give this sequence of poles 
        \begin{equation}
            \begin{split}
                & z_1 = (pq)^{1/4} v^{-1} u^{-(2n-1)/4} x_1 p^{k_1} q^{l_1}, \\
                & z_{i+1} = u z_i p^{k_i} q^{l_1}, \\
                & z_n = (pq)^{-1/4} v u^{(2n-1)/4} x_4^{-1} p^{-k_n} q^{-l_n}, \\
                & z_{n+1} = (pq)^{1/4} v^{-1} u^{-(2n-1)/4} x_2 p^{k_{n+1}} q^{l_{n+1}}, \\
                & z_{i+n+1} = u z_{i+n} p^{k_{i+n}} q^{l_{i+n}}, \\
                & z_{2n} = (pq)^{-1/4} v u^{(2n-1)/4} x_3^{-1} p^{-k_{2n}} q^{-l_{2n}}.
            \end{split}
        \end{equation}
    Along the lines of the works \cite{Gaiotto:2012xa, Spiridonov:2014cxa, Giacomelli:2023zkk, Bajeot:2023gyl} we can implement the Higgsing of the theory at the level of the SCI by taking the residues in the poles that pinch the contour of integration. Indeed, the sequence of poles reported above collide at $k_i = l_i = 0$, for $i=1,...,2n$, and, when the following identifications are operated, $x_1 = x_4^{-1}$, $x_2 = x_3^{-1}$ and $v = (pq)^{1/4} u^{-1/4}$, they pinch the contour of integration at (where we redefine $u \rightarrow t$)
    \begin{equation}
        \begin{split}
            & z_1 = x_1 t^{(1-n)/2}, \\
            & ... \\
            & z_n = x_1 t^{(n-1)/2}, \\
            & z_{n+1} = x_2 t^{(1-n)/2}, \\
            & ... \\
            & z_{2n} = x_2 t^{(n-1)/2}. \\
        \end{split}
    \label{Eq:Poles_Pinching}
    \end{equation}
    We modify the contour of integration in order to pick up only one time the poles contained in \eqref{Eq:Poles_Pinching}, taking the residues in these poles and implementing the identifications written above, we can resolve the $2n-$dimensional integral, completely breaking the gauge group. The interpretation of this procedure is that the chiral operator $A^{n-1} q^2$ constructed from the combination of gamma function \eqref{Eq:Gamma_Combination} is taking a vev \footnote{Note that without the electric deformation $A^n Q^2$, the operator $A^{n-1} q^2$ would be charged under the global symmetries. The identifications on the fugacities $u,v, x_u$ are consistent with the symmetry breaking induced by the electric deformation and make $A^{n-1} q^2$ uncharged under the global symmetries.}. By taking the residues we indeed obtain 
        \begin{equation}
            I_M = \prod_{l=1}^{2n} \prod_{1 \leq a < b \leq 4} \Gamma (t^{n+1-l} s_a s_b) \prod_{k=0}^{n-1} \prod_{a=1}^4 \prod_{u=1}^2 \Gamma ((pq)^{\frac{1}{2}} t^{-\frac{1}{2}-k} s_a^{-1} x_u^{\pm 1}),
        \end{equation}
        which represents the correct magnetic WZ theory. The first product represents the fields $W_1^2 A^{j}$, $j=0,..., 2n-1$ and the second one contains the fields $W_1 Q_1 A^k$, $k=0,...,n-1$.

\paragraph{$\bullet \quad l_1 = 3$ and $n_1 = 1$ case:}
    We follow the very same steps we presented in the field theory approach. We deconfine the rank-$2$ anti-symmetric tensor, then we dualize the $USp(6)$ gauge group and we integrate the massive fields. These steps are done by using the integral identities collected in \cite{rains2012elliptic, Spiridonov:2009za} and the reflection equation for the elliptic gamma functions \eqref{Eq:Reflection_Eq}. We skip these standard elementary steps and we focus on the quiver described in Figure \ref{Q:SWV_el_dualized_l1=3_n1=1}. The $S^3 \times S^1$ supersymmetric index for this theory is given by the formula
        \begin{equation}
            \begin{split}
                & I_E = \frac{(p,p)_{\infty}^{4} (q,q)_{\infty}^{4}}{2^{4} 4!} \Gamma (t^3 uv) \prod_{u=1}^{3} \Gamma ((pq)^{1/2} t^{5/4} v x_{u}^{\pm 1}) \prod_{1 \leq u < v \leq 3} \Gamma ((pq) t^{-1} x_{u}^{\pm 1} x_v^{\pm 1})\Gamma (pq) t^{-1}) \\ 
                & \prod_{u=1}^3 \prod_{a=1}^3 \Gamma ((pq)^{1/2} t^{-3/4} s_a^{-1} x_u^{\pm 1}) \prod_{l=0}^1 \prod_{1 \leq a < b \leq 3} \Gamma ((pq) t^{\frac{1}{2} -l} s_a^{-1} s_b^{-1} ) \prod_{l=1}^2 \prod_{a=1}^3 \Gamma ((pq) t^{\frac{1}{2} - l} s_a^{-1} u^{-1})\\
                & \Gamma (t)^2 \int_{\mathbb{T}^{2}} \int_{\mathbb{T}^{2}} \prod_{i=1}^2 \frac{\text{d} z_i}{2 \pi z_i} \prod_{k=1}^2 \frac{\text{d} w_k}{2 \pi w_k} \frac{\prod_{1 \leq k < l \leq 2} \Gamma (t w_k^{\pm 1} w_l^{\pm 1})}{\prod_{k=1}^2 \Gamma (w_k^{\pm 2}) \prod_{1 \leq k < l \leq 2} \Gamma (w_k^{\pm 1} w_l^{\pm 1})} \prod_{k=1}^2 \Gamma (t^{-3/4} w_k^{\pm 1} u^{-1}) \\
                & \prod_{a=1}^3 \Gamma (t^{1/4} s_a^{-1} w_k^{\pm 1}) \prod_{i=1}^2 \Gamma ((pq)^{1/2} t^{-1/2} z_i^{\pm 1} w_k^{\pm 1}) \frac{1}{\prod_{i=1}^2 \Gamma (z_i^{\pm 2}) \prod_{1 \leq i < j \leq 2} \Gamma (z_i^{\pm 1} z_j^{\pm 1})} \\
                & \prod_{i=1}^2 \Gamma ((pq)^{1/2} t^{-7/4} z_i^{\pm 1} v^{-1}) \prod_{a=1}^3 \Gamma ((pq)^{1/2} t^{1/4} z_i^{\pm 1} s_a) \prod_{u=1}^3 \Gamma (t^{1/2} z_i^{\pm 1} x_u^{\pm 1}) .
            \end{split}
        \end{equation}
    In the previous formula we defined $s_a$, $a=1,2,3$, and $u$ as the fugacities of the broken $SU(4)$, $z_i$ of the dual gauge group $USp(4)$, $w_k$ of the auxiliary gauge group $USp(4)$, $x_u$ of the flavor group $USp(6)$, $v$ of the  fictitious $U(1)$ symmetry group used to deconfine\footnote{ Indeed  no dynamical field transforms under this symmetry in the infrared and the index is independent of its value.}  
        and $t$ for the Abelian $U(1)$ symmetry. The balancing conditions are $u \prod_{a=1}^3 s_a = 1$ and $u v = 1$. In this index we can isolate the following terms
        \begin{equation}
            \begin{split}
                & I = \frac{(p,p)_{\infty}^{2} (q,q)_{\infty}^{2}}{2^{2} 2!} \prod_{l=0}^1 \prod_{1 \leq a < b \leq 3} \Gamma ((pq) t^{1/2 -l} s_a^{-1} s_b^{-1} ) \prod_{l=1}^2 \prod_{a=1}^3 \Gamma ((pq) t^{1/2 - l} s_a^{-1} u^{-1}) \\
                & \Gamma ((pq)^{1/2} t^{1/4} z_i^{\pm 1} s_a) \Gamma (pq) t^{-1}) \Gamma (t)^2 \int_{\mathbb{T}^2} \prod_{k=1}^2 \frac{\text{d} w_k}{2 \pi w_k} \frac{\prod_{1 \leq k < l \leq 2} \Gamma (t w_k^{\pm 1} w_l^{\pm 1})}{\prod_{k=1}^2 \Gamma (w_k^{\pm 2}) \prod_{1 \leq k < l \leq 2} \Gamma (w_k^{\pm 1} w_l^{\pm 1})} \\
                & \prod_{k=1}^2 \Gamma (t^{-3/4} w_k^{\pm 1} u^{-1}) \prod_{a=1}^3 \Gamma (t^{1/4} s_a^{-1} w_k^{\pm 1}) \prod_{i=1}^2 \Gamma ((pq)^{1/2} t^{-1/2} z_i^{\pm 1} w_k^{\pm 1}),
            \end{split}
        \end{equation}
    which, after the re-scaling $u = t^{-7/4} \Tilde{u}$ and $s_a = t^{-3/4} \Tilde{s}_a$ (note that in the re-scaled variable the balancing condition $\Tilde{u} \prod_{a=1}^3 \Tilde{s}_a = t^4$ is the one we need to use the integral identity), has the form of the confining integral in the case $l_1=2$ and $n=1$. Using the confining identity and the reflection equation \eqref{Eq:Reflection_Eq}, we arrive at the following index, where we have restored the not re-scaled fugacities $u, s_a$,
        \begin{equation}
            \begin{split}
                I_M = & \frac{(p,p)_{\infty}^{2} (q,q)_{\infty}^{2}}{2^{2} 2!} \Gamma (t^3 uv) \prod_{u=1}^{3} \Gamma ((pq)^{1/2} t^{5/4} v x_{u}^{\pm 1}) \prod_{1 \leq u < v \leq 3} \Gamma ((pq) t^{-1} x_{u}^{\pm 1} x_v^{\pm 1}) \\
                & \prod_{u=1}^3 \prod_{a=1}^3 \Gamma ((pq)^{1/2} t^{-3/4} s_a^{-1} x_u^{\pm 1}) \int_{\mathbb{T}^2} \prod_{i=1}^2 \frac{\text{d}z_i}{2 \pi z_i} \frac{\Gamma ((pq)^{1/2} t^{-5/4} z_i^{\pm 1} u^{-1})}{\prod_{i=1}^2 \Gamma (z_i^{\pm 2}) \prod_{1 \leq i < j \leq 2} \Gamma (z_i^{\pm 1} z_j^{\pm 1})} \\
                & \Gamma ((pq)^{1/2} t^{-7/4} z_i^{\pm 1} v^{-1}) \Gamma (t^{1/2} z_i^{\pm 1} x_u^{\pm 1}).
            \end{split}
        \end{equation}
    Following the field theory procedure we implement the confining case of Intriligator-Pouliot duality. Indeed, with the identifications $t_i = t^{1/2} x_i$ and $t_{i+3} = t^{1/2} x_{i+3}^{-1}$, with $i=1,2,3$, $t_7 = (pq)^{1/2} t^{-7/4} v^{-1}$, $t_8 = (pq)^{1/2} t^{-5/4} u^{-1}$ and $u= v^{-1}$, we recognize the integral identity of the Intriligator-Pouliot confining case. Recalling the definition $s_4 = u$, we obtain the expected magnetic index for the flipped theory
        \begin{equation}
            I_M = \prod_{a=1}^4 \prod_{u=1}^3 \Gamma ((pq)^{1/2} t^{-3/4} s_a^{-1} x_u^{\pm 1}).
        \end{equation}
        
\paragraph{$\bullet \quad l_1 = l$ and $n_1 = 1$ case:}
    The proof of this case is completely analogue to the previous one, the interested reader can reproduce it by following the stepwise procedure that we described before.
\paragraph{$\bullet \quad l_1 = l$ and $n_1 = n$ case:}
    As before, let us focus only on the theory in Figure \ref{Q:SWV_el_dualized_l1=l_n1=n}, all the other steps follows from the standard integral identities we have already mentioned before. The index of this quiver is
        \begin{equation}
            \begin{split}
                & I_E = \frac{(p,p)_{\infty}^{nl+l-2} (q,q)_{\infty}^{nl+l-2}}{2^{nl+l-2} (nl+l-2)!} \Gamma (t^{nl} uv) \prod_{u=1}^{l} \Gamma ((pq)^{1/2} t^{\frac{3nl-2-2n}{4}} v x_{u}^{\pm 1}) \Gamma (pq) t^{-1}) \Gamma (t)^{nl-1} \\
                & \prod_{u=1}^l \prod_{a=1}^3 \Gamma ((pq)^{1/2} t^{- \frac{nl-2+2n}{4}} s_a^{-1} x_u^{\pm 1}) \prod_{l=0}^{nl-2} \prod_{1 \leq a < b \leq 3} \Gamma ((pq) t^{nl/2 -1 -l} s_a^{-1} s_b^{-1} ) \\
                & \prod_{l=1}^{nl-2} \prod_{a=1}^3 \Gamma ((pq) t^{nl/2 -1 -l} s_a^{-1} u^{-1}) \prod_{1 \leq u < v \leq l} \Gamma ((pq) t^{-n} x_{u}^{\pm 1} x_v^{\pm 1}) \int_{\mathbb{T}^{nl-1}} \int_{\mathbb{T}^{l-1}} \prod_{i=1}^{l-1} \frac{\text{d} z_i}{2 \pi z_i} \\
                & \prod_{k=1}^{nl-1} \frac{\text{d} w_k}{2 \pi w_k} \frac{\prod_{1 \leq k < l \leq nl-1} \Gamma (t w_k^{\pm 1} w_l^{\pm 1})}{\prod_{k=1}^{nl-1} \Gamma (w_k^{\pm 2}) \prod_{1 \leq k < l \leq nl-1} \Gamma (w_k^{\pm 1} w_l^{\pm 1})} \prod_{k=1}^{nl-1} \Gamma (t^{-nl/4} w_k^{\pm 1} u^{-1}) \\
                & \prod_{a=1}^3 \Gamma (t^{-\frac{nl-4}{4}} s_a^{-1} w_k^{\pm 1}) \prod_{u=1}^l \Gamma ((pq)^{1/2} t^{\frac{1-n}{2}} w_k^{\pm 1} x_u^{\pm 1}) \frac{\prod_{i=1}^{l-1} \Gamma ((pq)^{1/2} t^{\frac{2-3nl}{4}} z_i^{\pm 1} v^{-1})}{\prod_{i=1}^{l-1} \Gamma (z_i^{\pm 2}) \prod_{1 \leq i < j \leq l-1} \Gamma (z_i^{\pm 1} z_j^{\pm 1})}  \\
                & \prod_{i=1}^{l-1} \Gamma ((pq)^{1/2} t^{-1/2} z_i^{\pm 1} w_k^{\pm 1}) \prod_{a=1}^3 \Gamma ((pq)^{1/2} t^{\frac{nl-2}{4}} z_i^{\pm 1} s_a) \prod_{u=1}^{l} \Gamma (t^{1/2} z_i^{\pm 1} x_u^{\pm 1}).
            \end{split}
        \end{equation}
    As before, we can isolate the following terms:
        \begin{equation}
            \begin{split}
                & I = \frac{(p,p)_{\infty}^{nl-1} (q,q)_{\infty}^{nl-1}}{2^{nl-1} (nl-1)!} \prod_{l=0}^{nl-2} \prod_{1 \leq a < b \leq 3} \Gamma ((pq) t^{nl/2 -1 -l} s_a^{-1} s_b^{-1} ) \Gamma (pq) t^{-1}) \Gamma (t)^{nl-1} \\
                & \prod_{l=1}^{nl-2} \prod_{a=1}^3 \Gamma ((pq) t^{nl/2 -1 -l} s_a^{-1} u^{-1}) \int_{\mathbb{T}^{nl-1}} \prod_{k=1}^{nl-1} \frac{\text{d} w_k}{2 \pi w_k} \frac{\prod_{1 \leq k < l \leq nl-1} \Gamma (t w_k^{\pm 1} w_l^{\pm 1})}{\prod_{k=1}^{nl-1} \Gamma (w_k^{\pm 2}) \prod_{1 \leq k < l \leq nl-1} \Gamma (w_k^{\pm 1} w_l^{\pm 1})} \\
                & \prod_{k=1}^{nl-1} \Gamma (t^{-nl/4} w_k^{\pm 1} u^{-1}) \prod_{a=1}^3 \Gamma (t^{-(nl-4)/4} s_a^{-1} w_k^{\pm 1}) \prod_{u=1}^l \Gamma ((pq)^{1/2} t^{(1-n)/2} w_k^{\pm 1} x_u^{\pm 1}) \\
                & \prod_{i=1}^{l-1} \Gamma ((pq)^{1/2} t^{-1/2} z_i^{\pm 1} w_k^{\pm 1}) ,
            \end{split}
        \end{equation}
    which, with the re-scaling $u=t^{-(M^{'}+5)/4} \Tilde{u}$ and $s_a = t^{- (M^{'}+1)/4} \Tilde{s}_a$, with $M^{'} = nl-1$, corresponds to the confining integral identity with $M^{'} = M-1$, $n_1^{'} = n_1 -1$, $l_1^{'} = l_1$ and $n_2^{'} = 1$, $l_2^{'} = l_1 -1$. The rest of the derivation is identical to the previous one, using the recursive index identity we arrive to the Intriligator-Pouliot confining theory, which finally gives the confined flipped theory. We skip these elementary steps. The last comment we make is that in the case of $n_1=2$ and $l_1=l$ the recursion lead to a theory with $n_1^{'} = n_2^{'}=1$ and $l_1^{'}=l_1$, $l_2^{'}=l-1$. It easy to see from equation \eqref{Eq:I_E}, that this former case is traced back to a theory with only one $USp(2(l_1^{'} + l_2^{'}))$ flavor group with $n_1= 1$. This feature allows us to implement the recursion at each step. The cases with more $USp(2l_i)$ flavor groups are completely analogue to the cases we have showed. This concludes the proof of the confining case from the supersymmetric index perspective.

\section{Duality}
\label{sec:dualcase}

    Let us study the duality case. We start with the same theory and same deformation as before, but the condition is that $M+N = \sum_{i=1}^K n_i l_i$ and we impose also the condition $M \geq N$. The field content and the general aspects of the duality have been discussed in section \ref{secduality}.
    The proof of the duality will be very similar to the confining case, so we will only emphasize the different steps.

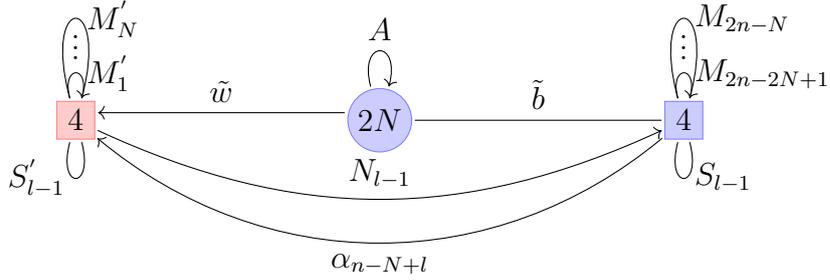
\begin{figure}[h]
    \centering
    \begin{tikzpicture}[shorten >=1pt,shorten <=1pt,node distance=4cm,on grid,auto]
    \node[rectangle, draw=red!50, fill=red!20, inner sep=2pt, minimum size=5mm] (1) at (-4,0) {$4$};
    \node[circle, draw=blue!50, fill=blue!20, inner sep=2pt, minimum size=5mm] (2) at (0,0) {$2N$};
    \node[rectangle, draw=blue!50, fill=blue!20, inner sep=2pt, minimum size=5mm] (3) at (4,0) {$4$};
    \node at (-4,1.07) {$\vdots$};
    \node at (4,1.07) {$\vdots$};
    \draw
    (1) edge[loop above] node[right] {$M_{1}^{'}$} (1)
    (1) edge[loop above, min distance=15mm] node[right] {$M^{'}_{N}$} (1)
    (1) edge[loop below, min distance=7mm] node[left] {$S^{'}_{l-1}$} (1)
    (3) edge[loop above] node[right] {$M_{2n-2N+1}$} (3)
    (3) edge[loop above, min distance=15mm] node[right] {$M_{2n-N}$} (3)
    (3) edge[loop below, min distance=7mm] node[right] {$S_{l-1}$} (3)
    (2) edge[loop above] node {$A$} (2);
    \draw[->]
    (2) edge[auto shift] node[above] {$\Tilde{w}$} (1)
    (1) edge[auto shift, bend right=25] node [above] {$N_{l-1}$} (3)
    (3) edge[auto shift, bend left = 40] node[below] {$\alpha_{n-N+l}$} (1);
    \draw (2) -- node {$\Tilde{b}$} (3);
    \node at (0,-3) {$W = \sum_{l=1}^N (M_{2n-2N+l} S_{N-l} + M_l^{'} S_{N-l}^{'} + \alpha_{n-N+l} N_{N-l} + S_l \Tilde{b}^2 A^{l} + S_l^{'} \Tilde{w}^2 A^{l} + $};
    \node at (0,-4) {$+ N_{l-1} \Tilde{w} \Tilde{b} A^{l}) + \sum_{\{l,j\}} M_l^{'} N_{j} N_{N+n+2-l-j}$};
    \end{tikzpicture}
    \caption{Quiver representation of the  electric confining theory $l_1 =2$ and $n_1=n$ after the consecutively use of dualities $(2.6)$ and $(2.9)$ of \cite{Razamat:2017hda}. In the Figure $l=1,...,N$ and in the term $M_l^{'} N_{j} N_{N+n+2-l-j}$ in the superpotential the indices $l,j$ take the values for which the fields $N$ and $M^{'}$ exist.}
    \label{Q:SWV_mag_l1=2_n1=n_dual}
    \end{figure}

\paragraph{$\bullet \quad l_1 = 2$ and $n_1 = n$ case:}
    In this case $M+N = 2n$, so we rewrite $M= 2n-N$ and we consider $M > N$ (when $M=N$ we only use (2.6) of \cite{Razamat:2017hda}). We flip all the would-be mesons $N_j$ and $M_{1,k_1}$. We consider the dual phase in Table (2.9) of \cite{Razamat:2017hda}, we depicted this theory in Figure \ref{Q:SWV_el_Higgsing_l1=2_n1=n_dual}.
    \begin{figure}[h]
    \centering
    \begin{tikzpicture}[shorten >=1pt,shorten <=1pt,node distance=4cm,on grid,auto]
        \node[rectangle, draw=red!50, fill=red!20, inner sep=2pt, minimum size=5mm] (1) at (-4,0) {$4$};
        \node[circle, draw=blue!50, fill=blue!20, inner sep=2pt, minimum size=5mm] (2) at (0,0) {$2(2n-N)$};
        \node[rectangle, draw=blue!50, fill=blue!20, inner sep=2pt, minimum size=5mm] (3) at (4,0) {$4$};
        \node at (-4,1.07) {$\vdots$};
        \node at (4,1.07) {$\vdots$};
        \draw
        (1) edge[loop above] node[right] {$M_1^{'}$} (1)
        (1) edge[loop above, min distance=15mm] node[right] {$M^{'}_{2n-N}$} (1)
        (1) edge[loop left] node[left] {$\beta_l$} (1)
        (2) edge[loop above] node {$A$} (2)
        (3) edge[loop above] node[right] {$M_1$} (3)
        (3) edge[loop above, min distance=15mm] node[right] {$M_{2n-M}$} (3);
        \draw[->]
        (1) edge[auto shift] node {$w$} (2)
        (3) edge[auto shift, bend left = 40] node[above] {$\alpha_l$} (1);
        \draw (2) -- node {$q$} (3);
        \node at (0,-2) {$W = M_{n-N} + \sum_{l=1}^{2n-N} M_l q^2 A^{l-1} + \sum_{l=1}^{2n-N} M_l^{'} w^2 A^{l-1} + \sum_{j=0}^{n-1} \alpha_j A^j w q $};
        \node at (0,-3) {$+ \sum_{k=0}^{2n-2N-1} \beta_k M_{2n-N-k}^{'}$};
    \end{tikzpicture}
    \caption{Quiver representation of the  electric confining theory $l_1 =2$ and $n_1=n$ after having turned on the superpotential deformation in the dual theory. In the Figure $l=0,...,n-1$.}
    \label{Q:SWV_el_Higgsing_l1=2_n1=n_dual}
    \end{figure}
    The duality dictionary is the one reported in equation \eqref{Eq:Duality_Mapping_72_Duals} and we see that some $M_l^{'}$ get mass, we integrate them out. Besides, we note in the superpotential in Figure \ref{Q:SWV_el_Higgsing_l1=2_n1=n_dual} that $M_{n-N}$ is linear and so $q^2 A^{n-1-N}$ takes vev. This vev breaks the $USp(2(2n-N))$ gauge group to $USp(2N)$ and the broken gauge group undergoes a color-flavor locking mechanism similar to the one presented in \cite{Bajeot:2023gyl}, such that new mesonic deformations of the type $A^j w q$, with $j=0,..., n-N-1$, are generated (and flipped by the corresponding $\alpha_j$). Furthermore, the color-flavor locking mechanism constructs the field $b$ from the off diagonal broken blocks of $A$. Expanding around the vev, one can verify that the fields $M_1,..., M_{2n-2N}$ become massive and that the following superpotential terms transforms as follows:
        \begin{equation}
            \begin{split}
                & \sum_{l= 2n-2N-1}^{2n-N} M_l q^2 A^{l-1} \longrightarrow  \sum_{l= 2n-2N-1}^{2n-N} M_l b^2 A^{l-(2n-2N+1)} \\
                & \sum_{l= 1}^{N} M_l^{'} w^2 A^{l-1} \longrightarrow \sum_{l= 1}^{N} M_l^{'} w^2 A^{l-1} + \sum_{i} M_i^{'} w^2 b^2 A^{N-n-2+i},
            \end{split}
        \end{equation}
    where in the second line the second sum is to be taken for the values of $i$ for which the power of $A$ is non-negative and the $M_i^{'}$ exist. So the theory we found after the Higgsing is reported in Figure \ref{Q:SWV_el_Higgsed_l1=2_n1=n_dual}.
    \begin{figure}[h]
    \centering
    \begin{tikzpicture}[shorten >=1pt,shorten <=1pt,node distance=4cm,on grid,auto]
        \node[rectangle, draw=red!50, fill=red!20, inner sep=2pt, minimum size=5mm] (1) at (-4,0) {$4$};
        \node[circle, draw=blue!50, fill=blue!20, inner sep=2pt, minimum size=5mm] (2) at (0,0) {$2N$};
        \node[rectangle, draw=blue!50, fill=blue!20, inner sep=2pt, minimum size=5mm] (3) at (4,0) {$4$};
        \node at (-4,1.07) {$\vdots$};
        \node at (4,1.07) {$\vdots$};
        \draw
        (2) edge[loop above] node {$A$} (2)
        (1) edge[loop above] node[right] {$M_1^{'}$} (1)
        (1) edge[loop above, min distance=15mm] node[right] {$M_{M}^{'}$} (1)
        (3) edge[loop above] node[right] {$M_{2n-2N+1}$} (3)
        (3) edge[loop above, min distance=15mm] node[right] {$M_{2n-N}$} (3);
        \draw[->]
        (1) edge[auto shift] node[above] {$w$} (2)
        (3) edge[auto shift, bend left = 30] node[above] {$\alpha_j$} (1);
        \draw (2) -- node {$b$} (3);
        \node at (0,-2) {$W = \sum_{l= 2n-2N-1}^{2n-N} M_l b^2 A^{l-(2n-2N+1)} + \sum_{l= 1}^{N} M_l^{'} w^2 A^{l-1} + \sum_{i} M_i^{'} w^2 b^2 A^{N-n-2+i}$};
        \node at (0,-3) {$+ \sum_{j=n-N}^{n-1} \alpha_j w b A^{N-n+j}$};
    \end{tikzpicture}
    \caption{Quiver representation of the  electric confining theory $l_1 =2$ and $n_1=n$ after the Higgsing. In the Figure $j=n-N,...,n-1$ and the sum over $i$ in the superpotential has to be taken for the values written in the text.}
    \label{Q:SWV_el_Higgsed_l1=2_n1=n_dual}
    \end{figure}
    Using consecutively the dualities in Table $(2.6)$ and $(2.9)$ of \cite{Razamat:2017hda} we obtain the  theory in Figure \ref{Q:SWV_mag_l1=2_n1=n_dual} (note that using these dualities the charges of $w$ and $b$ are exchanged). Integrating out the massive fields we obtain $W=0$ and the identifications are the following
        \begin{equation}
            \begin{split}
                & \Tilde{w} \longleftrightarrow w_1, \\
                & \Tilde{b} \longleftrightarrow q_1, \\
                & A \longleftrightarrow Y. \\
            \end{split}
        \end{equation}
    So, we are left with the magnetic theory we expected from the integral identity.
Observe that, the fact that the dual superpotential appearing in Figure \ref{Q:SWV_mag_l1=2_n1=n_dual} vanishes after integrating out the massive fields, is consistent with the flip of the mesons $N_j$ and $M_{1,k_1}$ in the electric theory. Furthermore the same comment will apply in the rest of the cases analyzed in this section.

    
\begin{figure}[h]
    \centering
    \begin{tikzpicture}[shorten >=1pt,shorten <=1pt,node distance=4cm,on grid,auto]
    \node[rectangle, draw=red!50, fill=red!20, inner sep=2pt, minimum size=5mm] (1) at (-4,0) {$3$};
    \node[rectangle, draw=red!50, fill=red!20, inner sep=2pt, minimum size=5mm] (4) at (-4,3) {$1$};
    \node[circle, draw=blue!50, fill=blue!20, inner sep=2pt, minimum size=5mm] (2) at (3.5,0) {$2l-2$};
    \node[rectangle, draw=blue!50, fill=blue!20, inner sep=2pt, minimum size=5mm] (3) at (6.5,0) {$2l$};
    \node[circle, draw=blue!50, fill=blue!20, inner sep=2pt, minimum size=5mm] (5) at (0,0) {$2N$};
    \node[rectangle, draw=red!50, fill=red!20, inner sep=2pt, minimum size=5mm] (6) at (-7,3) {$1$};
    \draw
    (5) edge[loop above] node[above] {$Y$} (5)
    (1) edge[loop left] node {$\alpha$} (1)
    (3) edge[loop below] node {$B_2$} (3);
    \draw [->] (5) edge[auto shift] node [above] {$m_1$} (1);
    \draw [->] (1) edge[auto shift, color = green] node [above] {$M_2$} (6);
    \draw (2) -- node [above]{$\Tilde{b}$} (5);
    \draw -- (2) edge[bend right= 60] node {$d$} (6);
    \draw -- (3) edge[bend right= 70, color = green] node [above]{$M_6$} (6);
    \draw (2) -- node [above]{$q_1$} (3);
    \draw [->] (3) edge[auto shift, bend right = 30] node [above]{$\gamma$} (4);
    \draw (5) -- node [right]{$f$} (4);
    \draw [->] (4) edge[auto shift, bend left = 10, color = green] node{$M_{2,0}^{''}$} (2);
    \draw  -- (4) edge[auto shift] node [above]{$E$}(6);
    \node at (0,-1.5) {$W = \gamma M_{2,0}^{''} q_1 + \Tilde{b}^2 q_1^2 + Y \tilde{b}^2 + d M_{2,0}^{''} E + \tau \text{Tr} (Y) + \alpha m_1^2 + \gamma M_{2,0}{''} q_1 + M_2 d m_1 \Tilde{b} +$};
    \node at (0, -2.5) {$ + M_6 d q_1 + B_2 q_1^2 + M_{2,0}^{''} f \tilde{b}$};
    \end{tikzpicture}
    \caption{Quiver representation of the  electric confining theory $l_1 =l$ and $n_1=1$ after the  duality on the $USp(2(l-N-1))$ gauge node. In this step we have redefined $\alpha_{l-2N-1} = \alpha$ and $\gamma_{l-2N-1} = \gamma$.}
    \label{Q:SWV_dual_l1=l_n1=1_Dual}
    \end{figure}
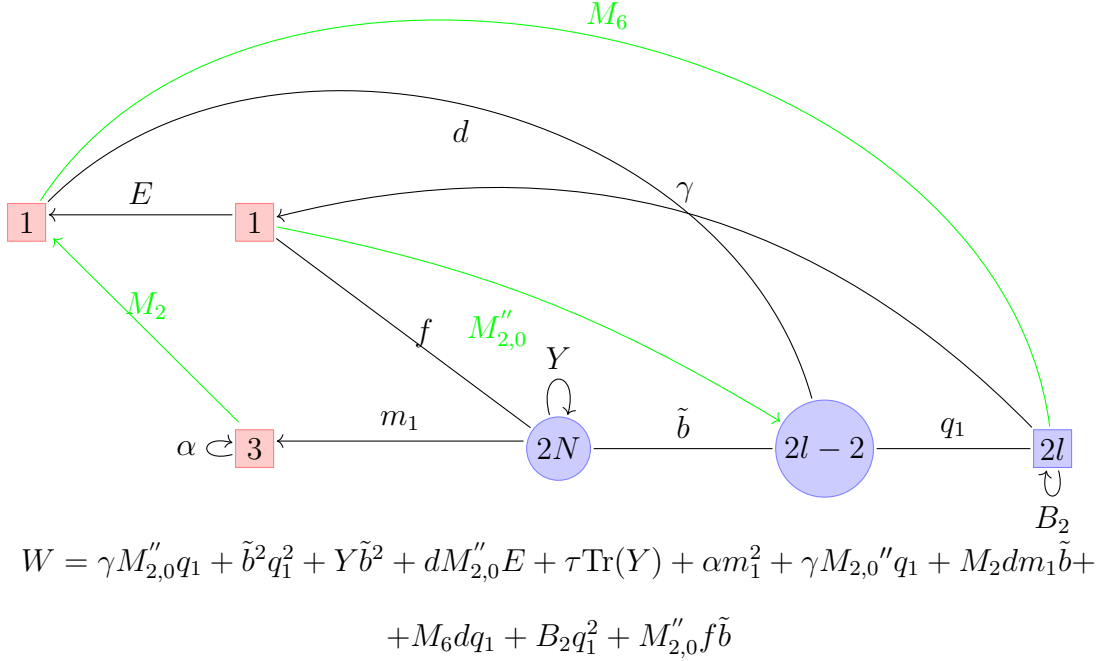

\paragraph{$\bullet \quad l_1 = l$ and $n_1 = 1$ case:}
    The procedure is the one we mentioned in the previous sections, so we do not go into too much details. After breaking the flavor symmetry, flipping all the would-be mesons, deconfining the anti-symmetric tensor, dualizing the original gauge group and integrating out the massive fields, we obtain a theory which is traced back to the case with $n_1^{'}=1$, $l_1^{'}=l-1$. Using the recursive step and integrating out the massive fields, we obtain the theory depicted in Figure \ref{Q:SWV_dual_l1=l_n1=1_Dual}.
    We dualize again the $Usp(2l-2)$ gauge node, it has $2l+2N+2$ fundamentals, so it becomes a $USp(2N)$ gauge node. Integrating out the massive fields (note that the  term $b^2 q_1^2$ becomes a quadratic meson, so it is massive), we can see that $USp(2N)$ has $2N+4$ fundamentals, so it confines. Confining this node we obtain the flipped magnetic theory expected from the index identity.

\paragraph{$\bullet \quad l_1 = l$ and $n_1 = n$ case:}
    The same steps listed above lead to a theory which incorporates the duality with $n_1{'}= n-1$, $l_1{'} = l$ and $n_2{'}=1$, $l_2^{'} = l-1$. In the following the procedure is the same and lead to the expected flipped magnetic dual. All the other cases are similar and can be analogously reproduced following the methodology we presented. The proof via the supersymmetric index follows the same steps we have described in the confining case, so we do not reproduce it here. The only difference is that in the first step with $l_1=2$ and $n_1=n$ the pinching poles do not completely eliminate the integral, but leave the expected $USp(2N)$ residual gauge group. All the other steps follows from the field theory and use the recursion and the already known integral identities.

\clearpage
\section{Conclusions}
\label{sec:conc}
In this paper we exploited an integral identity, conjectured in \cite{rains2012elliptic} and proven in \cite{vandebult2009elliptic}, in order to propose an infrared duality for $USp(2N)$ gauge theories with a rank-two index antisymmetric tensor and fundamental matter fields. The existence of such a duality deriving from the identity of \cite{rains2012elliptic}  was first conjectured in \cite{Spiridonov:2009za}, where the field content, the 
global symmetry structure and the dictionary of the duality were 
furnished. Here we have given a derivation of the duality in terms of other known dualities. The derivation given here has required various steps and techniques. These are associated to the combined and sequential action of dualities, tensor deconfinements, RG flows and Higgsing. 
This procedure has also been applied to the supersymmetric index, 
giving an alternative derivation of the integral identity of \cite{rains2012elliptic} itself.

Despite the fact that  the duality  passes non-trivial 
checks, for example, beyond the matching of the superconformal index, we checked the 't Hooft anomaly matching, we have not discussed many issues. It would be interesting for example to study  
infrared behavior of the dual theories, the existence of a conformal window, the possible presence of accidental symmetries and the stability of the superpotential.

For example in the regime where both the theories are UV free we studied the existence of a conformal window, at least for $K=1$.
We have seen that for $n_1=1$ there is a region of $l$ and $M$ where the unitarity bound is not violated and where the claim 
of a duality in a conformal window is consistent. For larger values of $K$ and/or of $n_j$ the situation becomes more complicated, because there are violations of the unitarity bound that imply the existence of accidental symmetries. In these cases the problem can be cured by flipping some singlets hitting the bound of unitarity. The situation nevertheless requires a case by case analysis that we did not perform here and that deserves further investigations.

Another interesting aspect that one can investigate is the reduction of the duality to three dimensions. In this case 
there should exists a 3d version of the duality discussed here with a KK monopole superpotential turned on. Further dualities should then be obtained through real mass flows or other deformations and Higgsing. It should be then interesting to figure out the whole picture emerging in such 3d case.

\bibliographystyle{JHEP}
\bibliography{ref.bib}

\providecommand{\href}[2]{#2}\begingroup\raggedright\begin{thebibliography}{10}

\bibitem{Seiberg:1994pq}
N.~Seiberg, \emph{{Electric - magnetic duality in supersymmetric nonAbelian
  gauge theories}},
  \href{https://doi.org/10.1016/0550-3213(94)00023-8}{\emph{Nucl. Phys. B}
  {\bfseries 435} (1995) 129}
  [\href{https://arxiv.org/abs/hep-th/9411149}{{\ttfamily hep-th/9411149}}].

\bibitem{Intriligator:1995ne}
K.~A. Intriligator and P.~Pouliot, \emph{{Exact superpotentials, quantum vacua
  and duality in supersymmetric SP(N(c)) gauge theories}},
  \href{https://doi.org/10.1016/0370-2693(95)00618-U}{\emph{Phys. Lett. B}
  {\bfseries 353} (1995) 471}
  [\href{https://arxiv.org/abs/hep-th/9505006}{{\ttfamily hep-th/9505006}}].

\bibitem{Berkooz:1995km}
M.~Berkooz, \emph{{The Dual of supersymmetric SU(2k) with an antisymmetric
  tensor and composite dualities}},
  \href{https://doi.org/10.1016/0550-3213(95)00400-M}{\emph{Nucl. Phys. B}
  {\bfseries 452} (1995) 513}
  [\href{https://arxiv.org/abs/hep-th/9505067}{{\ttfamily hep-th/9505067}}].

\bibitem{Luty:1996cg}
M.~A. Luty, M.~Schmaltz and J.~Terning, \emph{{A Sequence of duals for Sp(2N)
  supersymmetric gauge theories with adjoint matter}},
  \href{https://doi.org/10.1103/PhysRevD.54.7815}{\emph{Phys. Rev. D}
  {\bfseries 54} (1996) 7815}
  [\href{https://arxiv.org/abs/hep-th/9603034}{{\ttfamily hep-th/9603034}}].

\bibitem{Benvenuti:2020gvy}
S.~Benvenuti, I.~Garozzo and G.~Lo~Monaco, \emph{{Sequential deconfinement in
  3d $ \mathcal{N} $ = 2 gauge theories}},
  \href{https://doi.org/10.1007/JHEP07(2021)191}{\emph{JHEP} {\bfseries 07}
  (2021) 191} [\href{https://arxiv.org/abs/2012.09773}{{\ttfamily
  2012.09773}}].

\bibitem{Benvenuti:2021nwt}
S.~Benvenuti and G.~Lo~Monaco, \emph{{A toolkit for ortho-symplectic
  dualities}}, \href{https://doi.org/10.1007/JHEP09(2023)002}{\emph{JHEP}
  {\bfseries 09} (2023) 002}
  [\href{https://arxiv.org/abs/2112.12154}{{\ttfamily 2112.12154}}].

\bibitem{Comi:2022aqo}
R.~Comi, C.~Hwang, F.~Marino, S.~Pasquetti and M.~Sacchi, \emph{{The SL(2,
  \ensuremath{\mathbb{Z}}) dualization algorithm at work}},
  \href{https://doi.org/10.1007/JHEP06(2023)119}{\emph{JHEP} {\bfseries 06}
  (2023) 119} [\href{https://arxiv.org/abs/2212.10571}{{\ttfamily
  2212.10571}}].

\bibitem{Bajeot:2022kwt}
S.~Bajeot and S.~Benvenuti, \emph{{S-confinements from deconfinements}},
  \href{https://doi.org/10.1007/JHEP11(2022)071}{\emph{JHEP} {\bfseries 11}
  (2022) 071} [\href{https://arxiv.org/abs/2201.11049}{{\ttfamily
  2201.11049}}].

\bibitem{Amariti:2022wae}
A.~Amariti and S.~Rota, \emph{{3d N=2 SO/USp adjoint SQCD: s-confinement and
  exact identities}},
  \href{https://doi.org/10.1016/j.nuclphysb.2022.116068}{\emph{Nucl. Phys. B}
  {\bfseries 987} (2023) 116068}
  [\href{https://arxiv.org/abs/2202.06885}{{\ttfamily 2202.06885}}].

\bibitem{Bottini:2022vpy}
L.~E. Bottini, C.~Hwang, S.~Pasquetti and M.~Sacchi, \emph{{Dualities from
  dualities: the sequential deconfinement technique}},
  \href{https://doi.org/10.1007/JHEP05(2022)069}{\emph{JHEP} {\bfseries 05}
  (2022) 069} [\href{https://arxiv.org/abs/2201.11090}{{\ttfamily
  2201.11090}}].

\bibitem{Bajeot:2022lah}
S.~Bajeot and S.~Benvenuti, \emph{{Sequential deconfinement and self-dualities
  in 4d$ \mathcal{N} $ = 1 gauge theories}},
  \href{https://doi.org/10.1007/JHEP10(2022)007}{\emph{JHEP} {\bfseries 10}
  (2022) 007} [\href{https://arxiv.org/abs/2206.11364}{{\ttfamily
  2206.11364}}].

\bibitem{Bajeot:2023gyl}
S.~Bajeot, S.~Benvenuti and M.~Sacchi, \emph{{S-confining gauge theories and
  supersymmetry enhancements}},
  \href{https://doi.org/10.1007/JHEP08(2023)042}{\emph{JHEP} {\bfseries 08}
  (2023) 042} [\href{https://arxiv.org/abs/2305.10274}{{\ttfamily
  2305.10274}}].

\bibitem{Amariti:2023wts}
A.~Amariti, F.~Mantegazza and D.~Morgante, \emph{{Sporadic dualities from
  tensor deconfinement}},  \href{https://arxiv.org/abs/2307.14146}{{\ttfamily
  2307.14146}}.

\bibitem{Csaki:1996sm}
C.~Csaki, M.~Schmaltz and W.~Skiba, \emph{{A Systematic approach to confinement
  in N=1 supersymmetric gauge theories}},
  \href{https://doi.org/10.1103/PhysRevLett.78.799}{\emph{Phys. Rev. Lett.}
  {\bfseries 78} (1997) 799}
  [\href{https://arxiv.org/abs/hep-th/9610139}{{\ttfamily hep-th/9610139}}].

\bibitem{Csaki:1996zb}
C.~Csaki, M.~Schmaltz and W.~Skiba, \emph{{Confinement in N=1 SUSY gauge
  theories and model building tools}},
  \href{https://doi.org/10.1103/PhysRevD.55.7840}{\emph{Phys. Rev. D}
  {\bfseries 55} (1997) 7840}
  [\href{https://arxiv.org/abs/hep-th/9612207}{{\ttfamily hep-th/9612207}}].

\bibitem{rains2012elliptic}
E.~M. Rains, \emph{Elliptic littlewood identities},  2012.

\bibitem{vandebult2009elliptic}
F.~J. van~de Bult, \emph{An elliptic hypergeometric beta integral
  transformation},  2009.

\bibitem{Assel:2015nca}
B.~Assel, D.~Cassani, L.~Di~Pietro, Z.~Komargodski, J.~Lorenzen and
  D.~Martelli, \emph{{The Casimir Energy in Curved Space and its Supersymmetric
  Counterpart}}, \href{https://doi.org/10.1007/JHEP07(2015)043}{\emph{JHEP}
  {\bfseries 07} (2015) 043}
  [\href{https://arxiv.org/abs/1503.05537}{{\ttfamily 1503.05537}}].

\bibitem{Kinney:2005ej}
J.~Kinney, J.~M. Maldacena, S.~Minwalla and S.~Raju, \emph{{An Index for 4
  dimensional super conformal theories}},
  \href{https://doi.org/10.1007/s00220-007-0258-7}{\emph{Commun. Math. Phys.}
  {\bfseries 275} (2007) 209}
  [\href{https://arxiv.org/abs/hep-th/0510251}{{\ttfamily hep-th/0510251}}].

\bibitem{Romelsberger:2005eg}
C.~Romelsberger, \emph{{Counting chiral primaries in N = 1, d=4 superconformal
  field theories}},
  \href{https://doi.org/10.1016/j.nuclphysb.2006.03.037}{\emph{Nucl. Phys. B}
  {\bfseries 747} (2006) 329}
  [\href{https://arxiv.org/abs/hep-th/0510060}{{\ttfamily hep-th/0510060}}].

\bibitem{Spiridonov:2009za}
V.~P. Spiridonov and G.~S. Vartanov, \emph{{Elliptic Hypergeometry of
  Supersymmetric Dualities}},
  \href{https://doi.org/10.1007/s00220-011-1218-9}{\emph{Commun. Math. Phys.}
  {\bfseries 304} (2011) 797}
  [\href{https://arxiv.org/abs/0910.5944}{{\ttfamily 0910.5944}}].

\bibitem{Dolan:2008qi}
F.~A. Dolan and H.~Osborn, \emph{{Applications of the Superconformal Index for
  Protected Operators and q-Hypergeometric Identities to N=1 Dual Theories}},
  \href{https://doi.org/10.1016/j.nuclphysb.2009.01.028}{\emph{Nucl. Phys. B}
  {\bfseries 818} (2009) 137}
  [\href{https://arxiv.org/abs/0801.4947}{{\ttfamily 0801.4947}}].

\bibitem{Razamat:2017hda}
S.~S. Razamat and G.~Zafrir, \emph{{E$_{8}$ orbits of IR dualities}},
  \href{https://doi.org/10.1007/JHEP11(2017)115}{\emph{JHEP} {\bfseries 11}
  (2017) 115} [\href{https://arxiv.org/abs/1709.06106}{{\ttfamily
  1709.06106}}].

\bibitem{Gaiotto:2012xa}
D.~Gaiotto, L.~Rastelli and S.~S. Razamat, \emph{{Bootstrapping the
  superconformal index with surface defects}},
  \href{https://doi.org/10.1007/JHEP01(2013)022}{\emph{JHEP} {\bfseries 01}
  (2013) 022} [\href{https://arxiv.org/abs/1207.3577}{{\ttfamily 1207.3577}}].

\bibitem{Spiridonov:2014cxa}
V.~P. Spiridonov and G.~S. Vartanov, \emph{{Vanishing superconformal indices
  and the chiral symmetry breaking}},
  \href{https://doi.org/10.1007/JHEP06(2014)062}{\emph{JHEP} {\bfseries 06}
  (2014) 062} [\href{https://arxiv.org/abs/1402.2312}{{\ttfamily 1402.2312}}].

\bibitem{Giacomelli:2023zkk}
S.~Giacomelli, C.~Hwang, F.~Marino, S.~Pasquetti and M.~Sacchi, \emph{{Probing
  bad theories with the dualization algorithm I}},
  \href{https://arxiv.org/abs/2309.05326}{{\ttfamily 2309.05326}}.

\end{thebibliography}\endgroup
\end{document}